\documentclass[onecolumn,superscriptaddress,preprint,showpacs,amsmath,amssymb]{revtex4-1}
\pdfoutput=1
\usepackage{booktabs,graphicx,pstricks,verbatim,url}
\usepackage{graphicx}% Include figure files
\usepackage{dcolumn}% Align table columns on decimal point
\usepackage{bm}% bold math
\usepackage{slashed}

\def\eeq{\end{equation}}
\def\beq{\begin{equation}}
\def\eeq{\end{equation}}
\def\ba{\begin{eqnarray}}
\def\aa{\end{eqnarray}}
\newcommand{\met}{\slashed {E}_{T}}

\newcommand{\be}{\begin{eqnarray}}
\newcommand{\ee}{\end{eqnarray}}
\newcommand{\lsim}{\stackrel{<}{\sim}}
\newcommand{\gsim}{\stackrel{>}{\sim}}

\newcommand{\gev}{{\rm GeV}}

\newcommand{\vev}[1]{{\langle #1 \rangle}} % discussion comments
\def\eg{{\it e.g.}}

\newcommand{\Afbtt}{\ensuremath{A_{\mathrm{FB}}^{t\bar{t}}}}

\newcommand{\mzp}{\ensuremath{M_{Z'}}}
\def\ltap{\ \raise.3ex\hbox{$<$\kern-.75em\lower1ex\hbox{$\sim$}}\ }
\def\gtap{\ \raise.3ex\hbox{$>$\kern-.75em\lower1ex\hbox{$\sim$}}\ }
\def\lsim{\ \raise.3ex\hbox{$<$\kern-.75em\lower1ex\hbox{$\sim$}}\ }
\def\gsim{\ \raise.3ex\hbox{$>$\kern-.75em\lower1ex\hbox{$\sim$}}\ }
\def\bea{\begin{eqnarray}}
\def\eea{\end{eqnarray}}
\begin{document}
\input{epsf}
\preprint{FERMILAB-PUB-11-179-T}

\title{An Effective $Z'$}

\author{Patrick J. Fox} 
\affiliation{Theoretical Physics Department,Fermi National Accelerator Laboratory, Batavia, Illinois, USA}

\author{Jia Liu}
\affiliation{Center for Cosmology and Particle Physics, Department of Physics, New York 
University, New York, NY}

\author{David Tucker-Smith}
\affiliation{Center for Cosmology and Particle Physics, Department of Physics, New York 
University, New York, NY}
\affiliation{Department of Physics, Williams College, Williamstown, MA 01267}
\affiliation{School of Natural Sciences, Institute for Advanced Study, Einstein Drive, Princeton, NJ 08540}

\author{Neal Weiner}
\affiliation{Center for Cosmology and Particle Physics, Department of Physics, New York 
University, New York, NY}
\affiliation{School of Natural Sciences, Institute for Advanced Study, Einstein Drive, Princeton, NJ 08540}

\date{\today}
%%%%%%%%%%%%
\begin{abstract}
We describe a method to couple  $Z'$ gauge bosons to the standard model (SM), without charging the SM fields under the $U(1)'$, but instead through effective higher dimension operators.  This method allows complete control over the tree-level couplings of the $Z'$ and does not require altering the structure of any of the SM couplings, nor does it contain anomalies or require introduction of fields in non-standard SM representations.  Moreover, such interactions arise from simple renormalizable extensions of the SM - the addition of vector-like matter that mixes with SM fermions when the $U(1)'$ is broken.  We apply effective $Z'$ models as explanations of various recent anomalies: the D0 same-sign dimuon asymmetry, the CDF $W+$ di-jet excess and the CDF top forward-backward asymmetry.  In the case of the $W+$ di-jet excess we also discuss several complementary analyses that may shed light on the nature of the discrepancy. We consider the possibility of non-Abelian groups, and  discuss implications for the phenomenology of dark matter as well.
\end{abstract}
%%%%%%%%%%%%

\maketitle

%%%%%%%%%%%%%%%%%%%%%%%%%%%%%%%%%%%%%%
\section{Introduction}
The Standard Model (SM) is without question one of the great triumphs of modern physics. The elements are simple: a renormalizable theory with chiral fermions and  spontaneously broken gauge symmetry. The absence of dangerous FCNCs, baryon- or lepton- number violating operators, or large EDMs is naturally realized simply by the assumption that the scale at which non-renormalizable operators are generated is suitably high, with the important exception of the QCD $\theta$ parameter. While the myriad small parameters in the SM are perplexing, and the origin of the repeated generations completely uninformed, it remains a beautiful theory, and in no small part because of the delicate cancellation of gauge anomalies between quarks and leptons.

In looking at possible scenarios for physics beyond the Standard Model (BSM), there are a variety of motivations. Aside from scenarios motivated by the hierarchy problem, such as supersymmetry, a natural extension would be simply to add new elements which are copies of those we already know, such as additional generations, or new gauge fields.

Additional chiral generations are tightly constrained \cite{Kribs:2007nz}, but vectorlike generations can be added at any scale, with the only concern generally being that at some point more matter will drive the couplings to a Landau pole at a low scale. If such fields are present at a low scale, they can be produced, and then detected through their decays.

A new gauge field - a $Z'$ - is also trivial to add. However, the question of how to detect it is much more complicated, because it will only be produced if SM fields actually couple to it. 
If we want to charge the  SM fermions  under a  $U(1)'$ without introducing anomalies, the only flavor-universal  option is for the charges to be proportional to $B-L$ (with  right-handed neutrinos included in the theory).  
There are stringent constraints from LEP~\cite{Alcaraz:2006mx} on  any fields that couple significantly to leptons, meaning that such a $Z'$ would have to be in the TeV range if its coupling strength is comparable to those of the SM electroweak interactions.
LEP constraints can be evaded by charging only the SM quarks under the $U(1)'$ and not the leptons, but this introduces anomalies if the charges are flavor universal, and these anomalies must be cancelled by additional heavy chiral fermions.

Assigning flavor-non-universal $U(1)'$ charges, or charging  only  left- or right-handed quarks or leptons, complicates the SM picture for fermion masses and mixings, either by forbidding renormalizable Yukawa couplings, or  by setting certain CKM elements to zero at tree level.  This ruins one of the nicest elements of the SM, which is that it can can account for the familiar properties and interactions  of quarks and leptons entirely through renormalizable couplings.     Is it possible to extend the theory to include an apparently exotic new gauge field -- a leptophobic $Z'$ for example, or one with flavor-violating couplings  --  without spoiling this feature?

As we shall explore, such an extension {\em is} possible by employing ``effective'' $Z'$s - $Z'$s that only pick up effective charges to SM fields through non-renormalizable operators. We will argue that simple UV realizations can be constructed, and that these scenarios offer rich possibilities for  $Z'$ phenomenology.

This paper is laid out as follows: first, in Section~\ref{sec:effectiveZ}, we introduce the possibility that there
is an energy regime in which  the effective theory  includes only the SM fields, a $U(1)'$ gauge field $Z'$, and the field(s) $\phi$ responsible for the $U(1)'$ breaking, with an effective Lagrangian of the form
\begin{equation}
{\mathcal L} = {\mathcal L}_{SM} + {\mathcal L_{Z', \phi}} + {\mathcal L_{\rm higher \; dim.}} - \lambda |H|^2 |\phi|^2.
\end{equation}
Here ${\mathcal L}_{SM}$ has the ordinary fields and couplings of the SM Lagrangian,  ${\mathcal L_{Z', \phi}}$ consists of terms involving $\phi$ and $Z'$ but no SM fields, and ${\mathcal L_{\rm higher \; dim.}}$ consists of non-renormalizable operators that couple $Z'$ and $\phi$ to  SM fields.  The idea is simply that the SM fields are not charged under the $U(1)'$ and so couple to the $Z'$ only through higher-dimension operators.     In this case we say that the theory has an effective $Z'$.   In  Section~\ref{sec:effectiveZ} we also present a  renormalizable UV completion that generates the $Z'$ couplings to SM fields.
As indicated above, the $\phi$ particle(s) can couple in a renormalizable fashion through the Higgs portal.  This coupling is interesting in its own right, but will not be the focus of our attention.  

In Section \ref{sec:apps} we discuss various phenomena that can be accommodated in effective $Z'$ models.  We show how the tree-level exchange of a flavor violating $Z'$ can lead to CP violation in $B_s-\bar{B}_s$ mixing and thus explain the D0 same-sign dimuon asymmetry~\cite{Abazov:2010hv}.  We illustrate how the feature in the dijet spectrum in the recent CDF~\cite{Aaltonen:2011mk} analysis of $Wjj$ events may be explained by $WZ'$ production.  However, we point out that $Z'$ explanations appear to be in slight tension with UA2 dijet searches, and also propose alternative channels that should be searched in at the Tevatron to help confirm, or deny, whether the excess is consistent with \emph{any} model involving $WX$ production, where $X$ decays to dijets.  The final example we use to demonstrate the $Z'$ setup is the top FB asymmetry measured at CDF~\cite{Aaltonen:2011kc}.  In this case there are non-trivial constraints on an effective $Z'$ explanation from measurements of the rate for single top, like-sign tops and the total $t\bar{t}$ production cross section.  In Section~\ref{sec:dm} we discuss the application of effective $Z'$s in the realm of dark matter (DM).  For instance if the coupling of DM to the SM is through $Z'$ exchange then the couplings to protons and neutrons become free parameters, as does the ratio of the spin-dependent to the spin-independent cross section.  Finally, in Section~\ref{sec:conclusions} we conclude.

%%%%%%%%%%%%%%%%%%%%%%%%%%%%%%%%%%%%%%
\section{An effective $Z'$}\label{sec:effectiveZ}

The common approaches~\cite{Langacker:2008yv} to adding a $Z'$ to the SM, \eg\ gauging $B-L$, convert Yukawa couplings into non-renormalizable operators, and require the addition of massive fermions, often with complicated charges, to cancel anomalies. However, there is another approach that avoids these complications.  
One simply adds the following operator to the SM,
\be
({M^{-2})^{i}_j}\bar q_i \gamma_\mu q^j \phi^* D^\mu \phi \supset({M^{-2})^{i}_j}\bar q_i \gamma_\mu q^j \phi^* Z^{\prime\mu} \phi~.
\label{eq:operator}
\ee
Here $\phi$ is a scalar field whose vev breaks the $U(1)'$, and $({M^{-2})^{i}_j}$ is a matrix of couplings with mass-dimension equal to $-2$. This operator  ``effectively'' charges the SM fields under the new gauge group, but anomaly cancellation is manifest within the effective theory, and the renormalizable couplings of the SM are preserved.

This prompts us to ask: are there any differences between this theory and one in which  we allow arbitrary charges, while deferring issues related to anomalies and Yukawas to a higher scale? We will address this question within a specific UV completion.

We begin with a toy model of a single generation of SM quarks $q$, uncharged under the $U(1)'$. We include a pair of quarks $Q$ and $Q^c$, where $Q$ has identical SM charges to the $q$, but also carries $U(1)'$ charge $+1$, while $Q^c$ is its vectorlike partner, canceling anomalies. We include the Lagrangian terms
\be
{\mathcal{L}}\supset -\mu Q Q^c - y \phi q Q^c.
\label{eq:hopterms}
\ee
The first term provides a mass for the $Q$ fields, while the second term, which we refer to as a $\phi$-kawa coupling (to distinguish from the SM Yukawas) generates a mass term between the SM quark fields and the heavy quarks. When the $U(1)'$ breaks, these terms provide a missing partner mechanism, such that the mass eigenstates are
\be
\tilde Q = \cos \theta Q + \sin \theta q \hskip 0.5in \tilde q = -\sin \theta Q + \cos \theta q,
\ee
where
\be
\sin \theta = \frac{y \vev{\phi}}{\sqrt{\mu^2+ y^2 \vev{\phi}^2}}
\label{eq:theta}
\ee
determines the mixing angle.  

The kinetic term for the field that mixes with $q$ is
\be
\bar Q \slashed{D} Q \supset g' \sin^2 \theta Z_\mu^\prime{ \bar{ \tilde q }}\gamma^\mu \tilde q.
\label{eq:kineticinsert}
\ee
Using (\ref{eq:theta}) and expanding in powers of $\langle \phi \rangle$, we recognize the leading term as the original operator of (\ref{eq:operator}) with $\phi$ set to its vev,  and we see the effective coupling of the $Z'$ is $g_{eff} = g' \sin^2\theta$. The mass of the heavy quark can be written as
\be
M_{\tilde Q} = \frac{y/\sqrt{2}}{  g' \sin \theta} M_{Z'} = \frac{y/\sqrt{2}}{  \sqrt{g' g_{eff}}} M_{Z'}, 
\label{eq:VQmass}
\ee
where we have used $M_{Z'} = \sqrt{2} g' \langle \phi \rangle$.

We can generalize this UV completion to involve multiple quarks $q_i$, but we instantly see three important elements that distinguish this from a usual $Z'$.

\begin{itemize}
\item{ The effective coupling is bounded from above by $g'$, but can otherwise take on any, even seemingly anomalous, value.}
\item{Since only one linear combination of $q_i$ enters into the expression in (\ref{eq:kineticinsert}), the rank of the matrix $({M^{-2})^{i}_j}$ is set by the number of heavy quarks $Q_i$.}
\item{Given the bound $g_{eff} < g'$, (\ref{eq:VQmass}) tells us that new quarks must appear in the theory at some scale below $\sim  4 \pi \mzp /g_{eff}$.}
\end{itemize}
The first observation makes intuitive sense, but is not obvious from (\ref{eq:operator}). The latter two are important predictions that allow one to explicitly test whether the SM fields genuinely carry additional charges, or only have ``effective'' charges in the low energy effective theory. 

Similar approaches have been explored previously. For example, \cite{Langacker:1988ur,Nardi:1992nq,Nardi:1994iv} considered charging new heavy fields in addition to the SM fields under a $Z'$, with the final couplings determined by the initial charges and mixing. The important difference here being that we do not charge the SM fields, and only couple through NR operators. This scenario can be motivated from more elaborate models, however. Extra dimensional theories have SM fields charged under many $Z'$s (the KK resonances). A fermion field in the bulk, with a different profile from the KK modes can couple to many of them. The effective $Z'$ scenario can be thought of as a ``deconstruction'' \cite{ArkaniHamed:2001nc} of this down to a one-site model. Regardless of the motivation, we shall see that this setup allows for a wide range of interesting phenomenology. 

\subsection{Flavor and New Gauge Interactions}

Since the effective charges of the SM fields are not dictated by anomaly cancellation, or even a sense of ``natural'' rational charge ratios, they can contribute a wide range of flavor violating observables - for better or worse. Indeed, if we assume no flavor structure and weak-scale suppression, such operators are strongly excluded by any number of observables. However, there is an approximate flavor symmetry of the SM, and so we should see whether such flavor concerns are forced upon us.

A simple examination of the effective theory shows that this is not the case. If we assume that the new physics respects the approximate $SU(3)^5$ flavor symmetry of the SM (which is broken by the Yukawas), then the leading operators are flavor preserving. The leading flavor violating $Z'$ couplings are 
\be
\bar f \lambda_f \lambda_f^\dagger \gamma_\mu f \phi^* D^\mu \phi %\bar q \lambda_d \lambda_d^\dagger \lambda_u  \not\!\! D u \phi^* D^\mu \phi \\
\ee
where $f=e^c,u^c,d^c,l$. For all but $l$, the diagonalization of the Yukawas will also diagonalize these terms, leaving no remaining FCNCs. For $l$, we expect a negligible piece arising proportional to the neutrino masses. 

\begin{figure}[h]
\centering
\includegraphics[width=0.4 \textwidth]{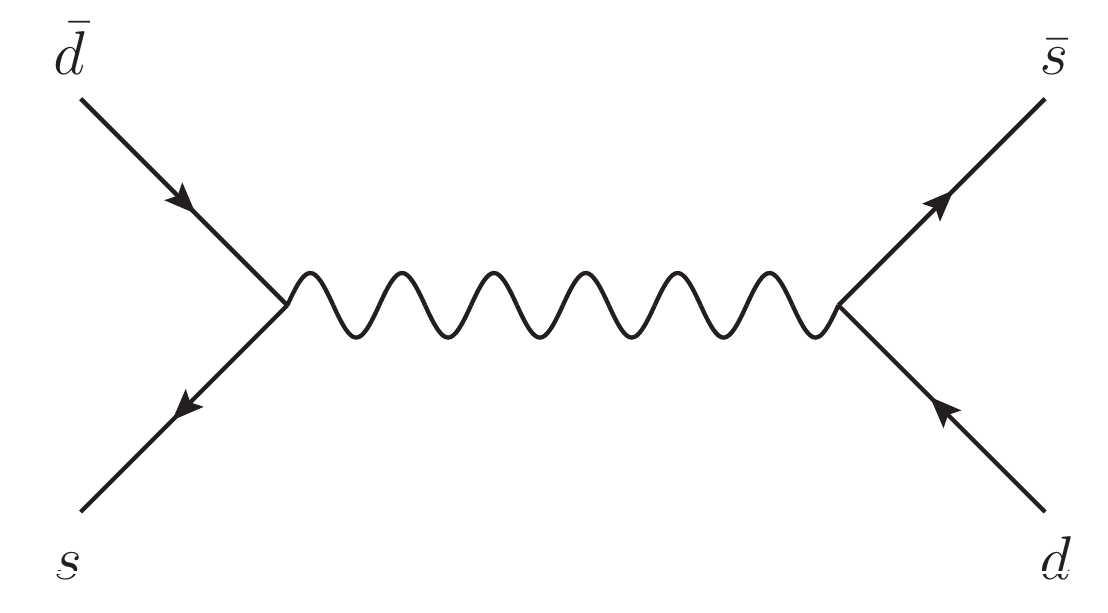}
\caption{A $Z'$ mediated FCNC.}\label{fig:fcncs}
\end{figure}

For $q$, the situation is somewhat different, as we must consider the operators
\be
\bar q( \lambda_u \lambda_u^\dagger+ \lambda_d \lambda_d^\dagger) \gamma_\mu q \phi^* D^\mu \phi \; .%\bar q \lambda_d \lambda_d^\dagger \lambda_u  \not\!\! D u \phi^* D^\mu \phi. \\
\ee
Diagonalizing the up components of $q$ leaves an operator
\be
\frac{1}{v^2}\bar u_LV_{CKM} M_d^2 V^{\dagger}_{CKM} \gamma_\mu u_L \phi^* D^\mu \phi %\bar q \lambda_d \lambda_d^\dagger \lambda_u  \not\!\! D u \phi^* D^\mu \phi \\
\ee
while for the down components, we have 
\be
\frac{1}{v^2}\bar d_L V^{\dagger}_{CKM} M_u^2 V_{CKM} \gamma_\mu d_L \phi^* D^\mu \phi %\bar q \lambda_d \lambda_d^\dagger \lambda_u  \not\!\! D u \phi^* D^\mu \phi \\
\ee
which can lead to dangerous contributions, such as to $\bar K - K$ mixing as in Figure \ref{fig:fcncs}. The mass and CKM suppressions are analogous to that for that of the usual GIM mechanism ($m_c^2$ vs $m_c^2 - m_s^2$), and so the coefficient of this operator must be small. However, if the coupling is generated at loop level, then Figure \ref{fig:fcncs} is effectively a two-loop process, making it easily safe compared to the SM contribution.
Thus, while flavor violating terms are possible, it is natural for them to be small.

We additionally must consider flavor {\em conserving} (but distinguishing) corrections, such as 
\be
\bar q \lambda_u \lambda_u^\dagger\gamma_\mu q h^* D^\mu h 
\label{eq:shiftzwcoupling}
\ee
which can be large for the third generation. We will discuss these terms in the context of specific scenarios.

\subsection{Kinetic Mixing}

At tree level effective $Z'$ models can be built to be hadrophobic, hadrophilic or neither, simply by the types of $\phi$-kawas that can be written down. At loop level, however, we generally expect the $Z'$ to couple to all SM fermions.  Consider, for instance, the situation where $\phi$-kawas are only allowed between SM quarks and heavy vector-like quarks, a hadrophilic model.  The vector-like quarks are charged under both hypercharge and the $U(1)'$ and so kinetic mixing of these two $U(1)$'s is induced, 
\bea
\mathcal{L}&\supset& -\frac{1}{4} Z_{\mu\nu}Z^{\mu\nu} -\frac{1}{4} A_{\mu\nu}A^{\mu\nu} -\frac{1}{4} b_{\mu\nu}b^{\mu\nu} + \frac{\chi}{2} b_{\mu\nu} (c_w A^{\mu\nu} - s_w Z^{\mu\nu}) \nonumber \\ 
&& -\frac{1}{2} \mzp^2 b_\mu b^\mu -\frac{1}{2} M_Z^2 Z_\mu Z^\mu
\eea
where $b$ is the $U(1)'$ gauge field, and we have worked in the mass basis for the SM photon and $Z$, after EW symmetry is broken.  The kinetic mixing term is marginal and receives contributions from physics at all scales above the mass of the particle in the loop.  All fields charged under both $U(1)$'s contribute, and in the unbroken theory the mixing coefficient is,
\be
\chi = \frac{g_Y g'}{16\pi^2} \mathrm{tr}\, Q_Y Q'\, \log \frac{\Lambda^2}{\mu^2}~.
\ee
Thus, the logarithmic divergence in $\chi$ is not present if there is a second pair of vector-like quarks, $\tilde{Q}$, $\tilde{Q}^c$, that have opposite charge to $Q$, $Q^c$ under $U(1)'$.  They need not mix with the SM fields.  As we discuss below, this situation naturally arises when the $U(1)'$ is embedded in a non-abelian group.  

By shifts in the photon ($A_\mu = \tilde{A}_\mu + c_w \chi b_\mu$) and $Z$  ($Z_\mu = \tilde{Z}_\mu -s_w \chi b_\mu$) we can remove the kinetic mixing terms, but the mass terms are no longer diagonal.  Finally, we can go to the mass eigenstate basis and determine the couplings of these mass eigenstates to the fermion currents $J_{em}$, $J_Z$ and $J_{Z'}$.  

Consider, for instance, the case in which the $Z'$ is leptophobic at tree level.
The $Z'-$lepton couplings generated by the loop-induced kinetic mixing  are contained in 
\be
\frac{e}{c_w} \chi Z'_\mu \left(c_w^2 J^\mu_{em}  - \frac{\mzp^2}{\mzp^2-M_Z^2} J^\mu_Z  \right)~,
\label{eq:kinmixingZprime}
\ee
where we work to leading order  in $\chi$, and where $J_Z^\mu = \sum_f  (T_3-Q s_w^2) \bar{f} \gamma^\mu f$ and $J_{em}^\mu= \sum_f Q \bar{f} \gamma^\mu  f$ are the gauge currents associated with $Z$ and $\gamma$ respectively. This $Z'$ has suppressed branching ratios to leptons but it is still possible to search for it in a di-lepton resonance at the Tevatron and LHC.

Similarly, kinetic mixing leads to shifts in the couplings between the SM $Z$ and any SM fields that mix with the vector-like matter charged under the $U(1)'$.  The modified couplings arise from the term
\be
g' \chi  \frac{s_w \mzp^2}{\mzp^2-M_Z^2} J_{Z'}^\mu Z_\mu \;, 
\label{eq:kinmixingZ}
\ee
once the fermion mixing has been taken into account and the SM component of $J_{Z'}$ has been identified.

\subsection{Non-Abelian Sectors and Isospin Violation}
\label{sec:nabiv}
Note that while we have focused on $U(1)'$ (Abelian) models, it is trivial to employ the same technique for non-Abelian models, so long as the structure of the fields breaking the $Z'$ group allow the presence of a $\phi$-kawa. For instance, we can consider an $SU(N)$ group, under which we have fields $Q$ and $\bar Q$ transforming as an $N$ and $\bar N$, and a field $\phi$ transforming as an $N$.  We can write $\phi q \bar Q$ just as before, which will yield the operator in (\ref{eq:operator}) precisely as before. As one enlarges the $Z'$ group, the number of fields contributing to SM $\beta$ functions increases, and if too much matter is added the theory will be driven  into a non-perturbative regime. However, the embedding into a larger group eliminates the presence of any kinetic mixing, as a non-Abelian group cannot mix renormalizably with a $U(1)$.
Only $S$-parameter type operators $F^Y_{\mu \nu} \phi^i \phi^*_j b_i^{j \mu\nu}/\mu^2$  are generated (with ${i,j}$ labeling $SU(N)$ gauge indices), and these should be small and thus safer from precision constraints.

This basic idea can be extended to more complicated scenarios. For instance,  consider the intriguing possibility of a gauged $SU(3)'$, with a set of fields $\phi_i^j$  transforming as  $(\bf 3, \bf \bar  3)$ under  $SU(3)'$  and the approximate (global) $SU(3)_{\rm flavor}$ of the SM. The $\phi$ vevs then break this down to the diagonal subgroup. The presence of $\phi_i^j q^i \bar Q_j$ (where $i$ labels the flavor group and $j$ the gauge group) then generate an effective gauging of flavor, again without fretting about anomalies.

Finally, we can even more exotic scenarios, such as ones involving isospin-violating couplings. Isospin violation in coupling to left-handed quarks is a challenge, because they are related by the SM $SU(2)$. One possibility would be to mix the neutral component of the $SU(2)$ gauge bosons with a $Z'$, although this would lead to tree-level corrections to electroweak precision operators, so any sizable coupling to SM fields would be challenging.

From the perspective of effective $Z'$s, these couplings can be achieved for instance through the operators
\be
c_d \bar q^i \gamma_\mu q_j \phi^* D^\mu \phi h^*_i h^j+c_u \bar q^i \gamma_\mu q_j \phi^* D^\mu \phi {h^*}^k h_l \epsilon_{ik} \epsilon^{jl}
\ee
where we assume that $\langle h_1 \rangle=0$ and $\langle h_2 \rangle=v$. These operators generate an isospin violating (distinguishing) but diagonal coupling to the $Z'$. For opposite signs of $c_u$ and $c_d$, these could give a scenario similar to what was conceived of in \cite{Buckley:2011vc}.

However, the operator is dimension {\em eight}, and would thus be expected to be very small. Moreover, it requires SM Higgs fields as well as $\phi$ vevs. To realize such an operator in the UV, one possible theory would be
\be
h q \bar U_0 + \mu_0 U_0 \bar U_0 + \mu_\pm U_+ \bar U_- + \lambda \phi U_0 \bar U_-
\ee
Here the subscript labels the $U(1)'$ charge. Diagonalizing the $U$ fields first and decoupling the heavier leads to an effective theory where the remaining $U$ field has an effective $Z'$ coupling, but still weds $q$ through a Yukawa coupling.
Invoking a similar term with $D$ would lead to $Z'$ interactions that would distinguish the up- and down- components of $q$. Note, however, that this requires not only significant field content, but also dramatic violation of custodial $SU(2)$ in the interactions of these new quarks. Loops involving these heavy fields would be expected to seriously impact precision electroweak observables. Invoking such a scenario without explicit UV realizations is essentially to sweep under the rug these issues which would very possibly exclude it.

%%%%%%%%%%%%%%%%%%%%%%%%%%%%%%%%%%%%%%
\section{Applications of effective $Z'$s}\label{sec:apps}

While effective $Z'$s can arise within simple models, it is not obvious what phenomenologies are consistent with our straightforward UV realizations.  In light of the recent anomalies reported from the Tevatron, we sketch out a few possibilities here, as we explore whether effective  $Z'$s can explain the D0 dimuon asymmetry, the top $A_{FB}$ anomaly, and the recently reported excess of events in $Wjj$ studies.

\subsection{Like-sign dimuon asymmetry}
\label{subsec:dimuons}

Using 6.1 fb$^{-1}$ of data, D0 has measured the like-sign dimuon asymmetry in semi-leptonic $b$-hadron decays,
\begin{equation} 
A^b_{sl} = \frac{N_b^{++}-N_b^{--}}{N_b^{++}+N_b^{--}}~,
\end{equation}
where $N_b^{++}$ and $N_b^{--}$ are the numbers of events with two semi-leptonically decaying $b$-hadrons that produce $\mu^+ \mu^+X$ and $\mu^- \mu^-X$, respectively \cite{Abazov:2010hv}.  The D0 result, $A^b_{sl} =(-9.57 \pm 2.51\pm 1.46)\times 10^{-3}$, differs from the SM prediction $A^b_{sl} ({\rm SM}) = (-2.0\pm 0.3)\times 10^{-4}$ \cite{Nierste:2011ti, Lenz:2006hd}, by 3.2$\sigma$.  Averaging the D0 result with a CDF measurement that used 1.6 fb$^{-1}$ of data, $A^b_{sl} =(8.0\pm 9.0\pm 6.8) \times 10^{-3}$ \cite{cdfAslnote}, gives $A^b_{sl} =(-8.5 \pm 2.8)\times 10^{-3}$, which still differs from the SM prediction by 2.9$\sigma$.  

Time-dependent tagged analyses of $B_s \rightarrow J/\psi \phi$ decays may also hint at new physics.  In the presence of a new phase $\phi_s^{\Delta}$ entering the $B^0_s - {\overline B}^0_s$ mass matrix, these analyses are sensitive to the phase $\phi_s^{\Delta}-2 \beta_s^{SM}$, where $\beta_s^{SM} =-\arg(-V_{ts}^* V_{tb}/V_{cs}^*V_{cb})$.  A combined CDF and D0 analysis using 2.8 fb$^{-1}$ of data for each experiment found $\phi_s^{\Delta}$ to deviate from zero at the $\sim 2\sigma$ level
\cite{cdfd0combinedBs}.
More recent CDF \cite{cdfBs2010} and D0 \cite{d0Bs2010}  studies using 5.2 fb$^{-1}$ and 6.1 fb$^{-1}$ of data, respectively, find smaller discrepancies, with the SM prediction lying near 68\% CL contours in the  $\Delta \Gamma_s -(\phi_s^{\Delta}-2 \beta_s^{SM})$ plane for both analyses.  

Finally, new physics may be responsible for the tension between the value of $V_{ub}$ as measured using  $B_d^{+} \rightarrow \tau^+ \nu$ decays and the value determined using unitarity triangle constraints.  In a global SM fit with ${\mathcal B}(B\rightarrow \tau \nu)$ not included in inputs, the predicted value of the branching ratio determined from the fit differs from the experimental value by 2.9 $\sigma$ \cite{Lenz:2010gu}.

The asymmetry $A^b_{sl}$ depends on the ``wrong charge'' asymmetries $a^s_{sl}$ and $a^d_{sl}$, where
\begin{equation}
a_{sl}^q = \frac{\Gamma({\overline B}^0_q(t) \rightarrow \mu^+X)-\Gamma(B^0_q(t) \rightarrow \mu^-X)}{\Gamma({\overline B}^0_q(t) \rightarrow \mu^+X)+\Gamma(B^0_q(t) \rightarrow \mu^-X)},
\end{equation}
and in principle the D0 like-sign dimuon asymmetry may be evidence for new physics affecting the $B_d$ system, the $B_s$ system, or both (no $b$ tags are required for the signal, so it is also possible that the new physics is not connected with these $B$-meson systems,  or the  D0 result may be  a statistical fluctuation and not a hint of new physics). The authors of \cite{Lenz:2010gu} carried out a global analysis of a scenario in which new physics is parametrized by complex parameters $\Delta_d$ and $\Delta_s$ that enter into the dispersive parts of the off-diagonal $B_d-{\overline B}_d$ and $B_s-{\overline B}_s$ mixing matrix elements, 
\begin{equation}
M_{12}^d = M_{12}^{SM,d} |\Delta_d| e^{i \phi^{\Delta}_d}  \quad \quad  \quad  \quad  \quad  \quad  
 M_{12}^s = M_{12}^{SM,s} |\Delta_s| e^{i \phi^{\Delta}_s}. 
\end{equation}
Because CP violation has been well measured in the $B_d$ system and is in general agreement with SM expectations, there is less room for a  large value of the phase $\phi^\Delta_d$ than for a large value of $\phi^\Delta_s$.  However,  the $B_d \rightarrow \tau \nu$ anomaly does prefer a non-zero value for $\phi_d^\Delta$, as  negative values of $\phi_d^\Delta$ increase the values of $\beta = \arg(-V_{td}^* V_{tb}/V_{cd}^*V_{cb})$ extracted from $B_d \rightarrow J/\psi K_S$ decays, which pushes $|V_{ub}|$ toward the larger values preferred by $B_d \rightarrow \tau \nu$.  A negative value for  $\phi_d^\Delta$ also gives a contribution to $A^b_{SL}$ that is in the right direction to address the same-sign dimuon anomaly.  The best fit point for the analysis of ref.~\cite{Lenz:2010gu} has a small but non-zero new phase in $B_d$ mixing, $\phi^\Delta_d = {-12.9^{+3.8}_{-2.7}}^\circ$ with $|\Delta_d| = 0.747^{+0.195}_{-0.082}$,  and a large new phase in $B_s$ mixing,  $\phi^\Delta_s = {-51.6^{+14.2}_{-9.7}}^\circ$ or  $\phi^\Delta_s = {-130^{+13}_{-12}}^\circ$  with $|\Delta_s| = 0.887^{+0.143}_{-0.064}$.  The SM hypothesis $\Delta_s = \Delta_d = 1$ is found to be disfavored at the 3.6$\sigma$ sigma level.  The analysis of ref. \cite{Lenz:2010gu} did not incorporate the most recent CDF and D0 $B_s \rightarrow J/\psi \phi$  analyses, which would lead to somewhat better agreement with the SM.   

The possibility of using flavor-violating $Z'$s to address $B$-physics anomalies been considered in \cite{Cheung:2006tm,Chiang:2006we,He:2009ie,Barger:2009eq,Barger:2009qs,Deshpande:2010hy, Alok:2010ij,Kim:2010gx}. 
To realize the effective $Z'$ idea in this context, we introduce a vector quark $(B,B^c)$ that is charged under a $U(1)'$.  The vector-quark mixes with the right-handed down-type quarks of the SM when a scalar field $\Phi$ acquires an vacuum expectation value $\langle \Phi \rangle = w$, which breaks the $U(1)'$.  The relevant Lagrangian terms for the fermion mixing are
\begin{equation}
{\mathcal L} \supset  -\sum_{i,j} \left[ \lambda_d^{ij} q_i d^c_j H+ y^{i} B d^c_i \Phi  + \mu  B B^c \right] + h.c.
\label{eqn:vqL}
\end{equation}
Working in the basis in which the SM Yukawa couplings are diagonal, but treating the associated quark masses as small perturbations compared to $\mu$, we see that the right-handed state that marries $B$ to become heavy is
\begin{equation}
{B^c}' \propto \mu B^c+ w \sum_i y^i  d^c_i.
\end{equation}
Similarly, the gauge eigenstate $B^c$ can be written in terms of mass eigenstates,
\begin{equation}
B^c = c_B {B^c}' + c_d d^c + c_s s^c+c_b b^c,
\end{equation}
which means that the $Z'$ inherits potentially flavor violating couplings to the light quarks:
\begin{equation}
g'  {\overline B^c} \overline{\sigma}^\mu B^c {Z'}_\mu \supset g' \sum_{i = d,s,b} (c_i^* c_j)\; {\overline d^c}_i  \overline{\sigma}^\mu d_j^c {Z'}_\mu. 
\end{equation}
Just as $B^c$ mixes with the right-handed quarks, $B$ also mixes with the left-handed quarks  at some level.  However, this mixing carries an extra Yukawa suppression beyond whatever suppression is present in the right-handed mixings.  This means that non-standard couplings of the light quarks to the $Z$ boson will be suppressed.  It also means that that the couplings of $Z'$ to the light left-handed quarks will be much smaller than those to the right-handed quarks, and so we neglect the left-handed couplings in our discussion.

Integrating out the $Z'$ and running down to the bottom quark mass generates effective Hamiltonian terms
\begin{equation}
{\mathcal H}_{eff}^{|\Delta B| = 2} = \frac{1}{2}\frac{{g'}^2}{M_{Z'}^2}  \eta^{6/23}  
\left[ 
(c_b^* c_s)^2 ({\overline b^c} \overline{\sigma}^\mu s^c)^2 +
(c_b^* c_d)^2 ({\overline b^c} \overline{\sigma}^\mu d^c )^2
 \right]+h.c.
\label{eqn:deltab=2}
\end{equation}
where $\eta = \alpha_s(M_{Z'})/ \alpha_s(m_b)$. In what follows we replace $\eta^{6/23}$ with ${\hat \eta}_B = 0.84$ \cite{Buras:1990fn,Lenz:2006hd}, effectively neglecting the running between $M_{Z'}$ and $M_{top}$.
These effective Hamiltonian terms contribute to the off-diagonal terms in the $B_s$-${\overline B}_s$ and  $B_d$-${\overline B}_d$  mass matrices,
\begin{equation}
M_{12}^{Z',q} = \langle B_q | {\mathcal H}_{eff}^{\Delta B = 2} | {\overline B}_q \rangle,
\label{eqn:m12}
\end{equation}
where $q=d,s$.
The relevant hadronic matrix element can be parametrized as
\begin{equation}
\langle B_q | ({\overline b^c} \overline{\sigma}^\mu q^c )^2| {\overline B}_q \rangle = \frac{1}{3} M_{B_q} f_{B_q}^2 B_{B_q},
\label{eqn:hadronicme}
\end{equation}
where the averages of lattice calculations given in   \cite{Lenz:2010gu} are $f_{B_s}\sqrt{B_{B_s}} = 212 \pm 13$ MeV and  $f_{B_d}\sqrt{B_{B_d}} = 174 \pm 12$ MeV. 

Using these equations we can determine what model-parameter values are preferred by the data. To illustrate we take the best-fit values of $\Delta_d$ and $\Delta_s$ from Ref.~\cite{Lenz:2010gu}.
The SM contributions to $M_{12}^{SM,q}$ are less precisely known than the measured mass splittings 
$\Delta M_s =17.77\pm0.012$ ps$^{-1}$ 
\cite{Abulencia:2006ze} and 
$\Delta M_d =0.507\pm 0.005$ ps$^{-1}$ 
\cite{Nakamura:2010zz}.
To an excellent approximation we have $\Delta M_q =  2 |M_{12}^q| = 2 |\Delta_q| |M_{12}^{SM,q}|$, and using this relation along with $M_{12}^q = M_{12}^{SM,q}+M_{12}^{Z',q}$, we derive 
\begin{equation}
M_{12}^{Z',q} =  \frac{\Delta M_q}{2} \left( \frac{\Delta_q-1}{ |\Delta_q|} \right) e^{i \phi_{M_q}^{SM}},
\label{eqn:deltagivesm12zprime}
\end{equation}
where  $\phi_{M_q}^{SM}$ is the phase of $M_{12}^{SM,q}$.
We can neglect $\phi_{M_s}^{SM}$ but  not $\phi_{M_d}^{SM}$, which comes from $(V_{td}^* V_{tb}) ^2 $. When new physics is allowed in $M_{12}^{d,s}$  the best-fit value found in \cite{Lenz:2010gu} is $\phi_{M_d}^{SM} = 55^\circ$.

Equations (\ref{eqn:deltab=2}), (\ref{eqn:m12}), (\ref{eqn:hadronicme}), and (\ref{eqn:deltagivesm12zprime}) can be used to to find
\begin{equation}
\frac{g' |c_b^* c_q|}{M_{Z'}} = 
\left[ \left(  \frac{3 \;\Delta M_q}{{\hat \eta}_B M_{B_q} f_{B_q}^2 B_{B_q}} \right)     \left|1-\frac{1}{\Delta_q}\right|   \right]^{1/2}, 
\end{equation}
which for the $B_d$ and $B_s$ systems gives
\begin{equation}
\frac{g' |c_b^* c_d|}{M_{Z'}} = 
 \left( \frac{0.84}{{\hat \eta}_B }\right)^{1/2}
  \left( \frac{174\;{\rm MeV}}{f_{B_d}\sqrt{ B_{B_d}}}\right)
\left(\frac{1}{3.7\times 10^2 \;{\rm TeV}} \right)  
 \left|1-\frac{1}{\Delta_d}\right|^{1/2} 
 \label{eqn:bdconstraint}
 \end{equation}
and
\begin{equation}
\frac{g' |c_b^* c_s|}{M_{Z'}} = 
 \left( \frac{0.84}{{\hat \eta}_B }\right)^{1/2}
  \left( \frac{212\;{\rm MeV}}{f_{B_s}\sqrt{ B_{B_s}}}\right)
\left(\frac{1}{76 \;{\rm TeV}} \right)  
 \left|1-\frac{1}{\Delta_s}\right|^{1/2}, 
 \label{eqn:bsconstraint}
\end{equation}
respectively.  For the inputs and  best-fit point of Ref.~\cite{Lenz:2010gu} these become
\begin{equation}
\frac{g' |c_b^* c_d|}{M_{Z'}} = 
(5.7\times 10^2 \;{\rm TeV})^{-1}
\label{eqn:bdbestfit}
\end{equation}
for the $B_d$ system, and
\begin{equation}
\frac{g' |c_b^* c_s|}{M_{Z'}} = 
\left\{
\begin{array}{c}
\left(55 \;{\rm TeV} \right)^{-1}  \quad\quad {\mathcal Re}(\Delta_s) <0 \\
  \left(79 \;{\rm TeV} \right)^{-1}\quad\quad {\mathcal Re}(\Delta_s) >0 
\end{array}
\right.
\label{eqn:bsbestfit}
\end{equation}
for the $B_s$ system.

Allowing for an order-one phase in its coefficient, the $|\Delta S| = 2$ operator
\begin{equation}
\frac{1}{\Lambda^2}  ({\overline d^c} \overline{\sigma}^\mu s^c)^2
\label{eqn:deltas=2}
\end{equation}
must be suppressed by a scale $\Lambda \ge 2 \times 10^4$ TeV to be consistent with the measured value of $\epsilon_K$ \cite{Isidori:2010kg}. Neglecting QCD corrections, integrating out the $Z'$ generates this operator with coefficient
\begin{equation}
\frac{1}{\Lambda^2}=\frac{g'^2  |c_s^* c_d|^2}{2 M_{Z'}^2} =\frac{1}{2}\left| \frac{c_d c_s}{c_b^2} \right| \left(  \frac{g' |c_b^* c_d|}{M_{Z'}} \right)\left(  \frac{g' |c_b^* c_s|}{M_{Z'}} \right)~.
\end{equation}
For the best-fit values of $\Delta_s$ and $\Delta_d$ we find
\begin{equation}
\Lambda=
\left\{
\begin{array}{c}
\sqrt{ \left| \frac{c_b^2}{c_d c_s} \right|}\times 2.5 \times 10^2 \;{\rm TeV}  \quad\quad {\mathcal Re}(\Delta_s) <0 \\
\sqrt{ \left| \frac{c_b^2}{c_d c_s} \right|}\times3.0 \times 10^2 \;{\rm TeV}  \quad\quad {\mathcal Re}(\Delta_s) >0 
\end{array}
\right. ,
\end{equation}
and the requirement $\Lambda \ge 2 \times 10^4$ TeV then corresponds to
\begin{equation}
\sqrt{ \left| \frac{c_b^2}{c_d c_s} \right|} \gsim
\left\{
\begin{array}{c}
80  \quad\quad {\mathcal Re}(\Delta_s) <0 \\
67  \quad\quad {\mathcal Re}(\Delta_s) >0
\end{array}
\right. 
\end{equation}

For example, suppose we take $g'=0.5$, $M_{Z'} = 150$ GeV, and $|c_b|= 0.6$.   Then Equations~(\ref{eqn:bdbestfit})  and (\ref{eqn:bsbestfit}) give $|c_d|= 8.8 \times 10^{-4}$ and  $|c_s|= 9.1 \times 10^{-3}$ or $ 6.3 \times 10^{-3}$, leading to  $\sqrt{|c_b^2/c_d c_s|} = 210$ or 250, consistent with the $|\epsilon_K|$ constraint.

Given that $|c_b|$ is much larger than $|c_s|$ and $|c_d|$ for this parameter point, the heavy vector quark $B^c$ mixes almost exclusively with $b^c$.  In the notation of section \ref{sec:effectiveZ} we thus identify $ |c_b| = \sin \theta = |y^bw|/\sqrt{\mu^2 + |y^b w|^2}$, where we neglect the bottom quark mass.  The mass of the heavy quark $B$ is $m_B = \sqrt{\mu^2+|y^b w|^2} = yw/\sin \theta$.  
Using the relation between the $Z'$ mass and  the $U(1)'$-breaking vev $w$,  $M_{Z'} = \sqrt{2} g' w$, we get 
\begin{equation}
m_B = \left(  \frac{|y^b|/\sqrt{2}}{g'}\right)  \frac{M_{Z'}}{\sin\theta}
= \left(  \frac{|y^b|/\sqrt{2}}{g'}\right) \frac{M_{Z'}}{|c_b|}.
\end{equation}
Taking the sample model parameters from above along with $|y^b|=1$, for example, we get $m_B = 350$ GeV.
For the same $\phi$-kawa, larger masses are possible by reducing $g'|c_b|$ while holding  $g'| c_b c_s|$ and $g' |c_b c_d|$ fixed to maintain the same effect on $B_d$ and $B_s$ mixing. In this way, there is room to increase the heavy quark mass by a factor of $\sim3$ while remaining consistent with the $\epsilon_K$ constraint.  
  Once produced, these $B$ quarks would likely decay through $B \rightarrow Z b \rightarrow b {\overline b} b$ or $B \rightarrow \phi b \rightarrow b {\overline b} b$.

The phases of the $c_q$ parameters are determined by the phases of the $\phi$-kawa terms, two of which cannot be removed by field redefinitions.  One of the field redefinitions that can be used to remove a phase is changing the phase of $B^c$, which shifts the phases of all the $c_q$ together.  Since the quantities of interest are always of the form $c_q^* c_{q'}$, none of these quantities can be set real through phase rotations.  If $c_q^* c_{q'}$ is complex due to complex $\phi$-kawa's then this phase is physical.

The phases of the $c_q$ are constrained through
\begin{equation}
\arg(c_b^* c_q)^2 = \arg\left[ (\Delta_q-1) e^{i \phi_{M_q}^{SM}}\right]~.
\end{equation}
The best-fit  point of \cite{Lenz:2010gu} gives
\begin{equation}
\arg(c_b^* c_d)^2 =
-93^\circ \quad\quad
\end{equation}
and
\begin{equation}
\arg(c_b^* c_s)^2 =
\left\{
\begin{array}{c}
-156^\circ  \quad\quad {\mathcal Re}(\Delta_s) <0 \\
-123^\circ   \quad\quad {\mathcal Re}(\Delta_s) >0
\end{array}
\right. .
\end{equation}
It follows that the phase of the coefficient of the $\Delta S = 2$ operator of Equation (\ref{eqn:deltas=2}) is
\begin{equation}
\arg(c_d^* c_s)^2=   \left\{
\begin{array}{c}
-64^\circ  \quad\quad {\mathcal Re}(\Delta_s) <0 \\
-30^\circ   \quad\quad {\mathcal Re}(\Delta_s) >0
\end{array}
\right. 
\end{equation}

Contributions to $\Delta F = 1$ operators are also generated by the $Z'$, but for the parameter point we have considered  these are much smaller than SM contributions.  An exception is the $({\bar b}^c {\bar \sigma}^\mu s^c)({\bar d}^c {\bar \sigma}_\mu s^c)$ operator, which  leads to the $\Delta B=1$, $\Delta S=2$ transition, $b\rightarrow ss \bar{d}$.  The constraint on this is comes from searches for the decay mode $B^-\rightarrow K^- K^-\pi^+$ whose branching ratio is bounded by $B(B^-\rightarrow K^- K^-\pi^+) < 1.6\times 10^{-7}$~\cite{Aubert:2008rr}.  In the SM this mode has both strong GIM and CKM suppression and has branching ratio less than $10^{-11}$~\cite{Fajfer:2001te}.  The example point described above predicts a rate comparable to that in the SM and so is far below present bounds.

The $\Delta B=1$ process $b\rightarrow s \gamma$ is generated at the loop level, through a loop involving a $b$ quark and a $Z'$.  For the example point above, this new contribution is a few percent correction to the SM amplitude and does not conflict with measurements.  

Kinetic mixing between $U(1)_Y$ and $U(1)'$ modifies the coupling of right-handed  $b$ quarks to the $Z$ boson and induces $Z'$ couplings to leptons.  From Eqn.~(\ref{eq:kinmixingZ}), we see that the shift in the coupling of $Z$ to  $b^c$ is 
\begin{equation}
\delta g_{b^c b^c Z} = \chi \; g'|c_b|^2   \frac{s_w \mzp^2}{\mzp^2-M_Z^2}, 
\end{equation}
which is a small correction if $\chi$ is generated at loop level without a large log.  The $Z'$ production cross section at LEP is also sufficiently small in this case.  For example, using Eqn.~(\ref{eq:kinmixingZprime}) and the sample point presented above, we use MadGraph to calculate the following cross section for $e^+ e^- \rightarrow Z' \gamma$ at $\sqrt{s} = 161$ GeV,
\begin{equation}
\sigma = \left(\frac{\chi}{10^{-2}} \right)^2\times 0.10 \; {\rm pb},
\end{equation}
compared to a total cross section into hadrons $(+ \;\gamma)$ of about 35 pb \cite{Alcaraz:2006mx}.
The process $B_s \rightarrow \ell^+ \ell^-$, can also occur at the one-loop level through loop-induced kinetic mixing, but because the process is helicity suppressed the constraint on $\chi$ is far weaker than those coming from LEP.

To prevent large kinetic mixing, the $U(1)'$ should be embedded in a non-abelian group such as $SU(2)$, with the vector quarks  filling out a complete multiplet of the non-abelian group.  Or, we can include a pair of vector quarks with opposite $U(1)'$ charge, so that their log-divergent contributions to the kinetic mixing cancel.

It is possible to supersymmetrize the  scenario we have described, in a manner consistent with gauge coupling unification. Instead of adding vector-quarks we add ${\bf 5} + {\bf {\bar 5}}$ pairs of chiral superfields, but only give sizable $\phi-$kawa couplings to the triplets so as to keep $Z'$ leptophobic.  We need to introduce two  ${\bf 5} + {\bf {\bar 5}}$ pairs to avoid a logarithmically divergent $\chi$, but the SM gauge couplings still remain perturbative up the the gut scale.  We also find that the $\phi-$kawa coupling $|y^b|$ remains perturbative to high scales for low-scale values as large as about 1, but with this matter content we find that we need $g' \ltap 1/3$ at low energies for it to remain perturbative.  This is smaller than the  $g' =1/2$ value adopted for the sample point above, but we could have chosen $g'$ to be smaller, compensated by larger values of $|c_d|$ and $|c_s|$ (subject to the $\epsilon_K$ constraint).  If the $U(1)'$ is embedded in an $SU(2)$, however, this issue is no longer a concern.  

In summary, an effective $Z'$   does quite well at addressing $b$-physics anomalies.  We have considered a minimal model in which a single vector-quark mixes with the right-handed down-type quarks of the SM.  If one only wants the new physics to enter the $B_s$ system, one can imagine that the only non-negligible $\phi-$kawas are to $s^c$ and $b^c$.  However, there may be a hint for new physics in the $B_d$ system as well, coming from $B \rightarrow \tau \nu$, and this can be addressed by having a small $\phi-$kawa with $d^c$ as well.  We have seen that it is possible to address the $b$-physics anomalies without coming into conflict with constraints from kaon physics.    For the viable parameter point we studied, with a $Z'$ mass of 150 GeV, and the vector quark is expected to be rather light, $m_Q \ltap 1$ TeV for $|y^b| = 1$.  Larger vector-quark masses are possible by increasing $|y^b|$ (while adjusting $\mu$ to hold the mixing angle fixed), although this eventually leads to a Landau  pole below the GUT scale.

In our discussion we have neglected the contributions of the $\phi$ particle to $B_s$ and $B_d$ mixing, and we now show that this was justified.  Neglecting  $m_d$ and $m_s$, working to first order in $m_b/\mu$, and using $|y^d w|,\;|y^s w| \ll \mu$, we find that the relevant flavor-violating couplings of $\phi$ are
\be
{\mathcal L} \supset \left( \frac{y^q y^b  m_b w}{\mu^2 + |y_b w|^2} \right) \; b q^c \frac{\phi}{\sqrt{2}} +h.c,
\ee
where $q=d, s$.  Using $c_b = (y^b w)/\sqrt{\mu^2 + |y_b w|^2}$ and $c_q = y^q w/\mu$ , we can express this term as
\be 
g'(c_b c_q)\frac{m_b \cos\theta}{M_{Z'}}\; b q^c\phi + h.c., 
\ee
giving four-fermion operators of the form
\be
\frac{1}{2}\frac{{g'}^2}{M_{Z'}^2}    
(c_b c_q)^2  \left( \frac{m_b \cos \theta}{ m_\phi} \right)^2 (b q^c)^2 +h.c.
\label{eq:phiexchange}
\ee
The relevant hadronic matrix elements in Equations (\ref{eqn:deltab=2}) and (\ref{eq:phiexchange}) are comparable, so we see that the contribution from $\phi$ exchange is suppressed by a factor $(m_b \cos\theta/m_\phi)^2$, which we would naively expect to be of order $10^{-3}$.  Contributions to $K-{\bar K}$ are even more suppressed relative to the $Z'$ contribution, as $m_b$ is replaced by $m_s$.  

\subsection{W+dijets signal}

In a recent CDF analysis which studied the invariant mass distribution of pairs of jets produced in association with a $W$, based on $\mathcal{L}_\ell = 4.3\, \mathrm{fb}^{-1}$ of data \cite{Aaltonen:2011mk}, an excess is observed, centered at $m_{jj} \approx 150$ GeV.  The excess is well fit, with a significance of $3.2\sigma$, by the addition of a gaussian component of width given by the expected dijet mass resolution at 150 GeV, and normalization corresponding to a production cross section of $\sigma_{Wjj}\approx 4$ pb.  Here we investigate the possibility that the excess is due to $W Z'\rightarrow l\nu jj$ events.  Related models, including theories involving $Z'$ bosons, have recently been proposed to explain the $Wjj$ excess~\cite{Eichten:2011sh,Buckley:2011vc,Yu:2011cw,Kilic:2011sr, AguilarSaavedra:2011zy,Cheung:2011zt,Nelson:2011us,Sato:2011ui,Wang:2011ta,He:2011ss,Anchordoqui:2011ag,
Dobrescu:2011px,Buckley:2011vs}.

Since we want a dijet signal in association with a $W$, the effective coupling must be to left-handed quarks, and the necessary operator is
\be
({M^{-2})^{i}_j} \bar q_{Li} \gamma_\mu q^j_L \phi^* D^\mu \phi ~.
\label{eq:lhopp}
\ee
Because this operator necessarily involves the down sector, as well, we must be careful to control flavor violating elements, thus, we assume that this operator is $U(3)$ flavor conserving, i.e., $({M^{-2})^{i}_j} \propto \delta^i_j$.

To realize this, we pursue the standard UV realization of the mixing of SM quarks with vector quarks charged under the $U(1)'$.  For the operator to respect the $U(3)$ flavor symmetry, we must introduce 3 vector-like pairs of quark doublets $Q_i,Q^c_i$ and mix them with the SM quarks in a flavor universal fashion.  The additional terms in the Lagrangian are
\be
\mathcal{L}\supset - (\mu Q^c_i Q_i + \lambda Q^c_i q_i \phi)~. 
\ee

Using MadGraph \cite{Alwall:2007st} with CTEQ6L1 parton distribution functions \cite{Pumplin:2002vw}, we find that  the production cross section for $WZ'$ is 29 pb, for a 150 GeV $Z'$ whose couplings to left-handed quarks are equal to 1.  For the actual cross section to be around 4 pb, we therefore need an effective coupling $g_{eff} = g' \sin^2\theta \sim 0.37$, where $g'$ is the $U(1)'$ gauge coupling and $\theta$ parametrizes the mixing between the left-handed SM quarks and the vector quarks.

Such a light $Z'$ is excluded, by Tevatron and others, if it has sizable couplings to leptons.  In the effective $Z'$ model with no massive vector-like leptons such couplings are forbidden at tree-level.  The leading constraint then comes from the UA2 search for a dijet resonance.  Dijet searches at UA2 at $\sqrt{s} = 630$ GeV \cite{Alitti:1993pn} constrain the production cross section for a 150 GeV $Z'$ that decays exclusively to $q \overline {q}$ to be below 121 pb at 90\% CL.   Tevatron dijet constraints are not competitive at this mass \cite{Aaltonen:2008dn}.  Again using CTEQ6L1 PDFs, and taking 
$Q^2 = \mzp^2$, we find that for a 150 GeV $Z'$ that couples only to left-handed quarks with coupling strength equal to one,  the leading-order cross section for resonant production at UA2 is 2,450 pb. This constrains $g_{eff}$ to be below about $0.23$, and thus the $WZ'$ cross section at the Tevatron to be below about 1.4 pb.  This is significantly less than the 4 pb estimate given in the CDF paper, but may be consistent with what's required for the $l\nu jj$ excess once all relevant uncertainties are taken into account.

If we allow the $Z'$ to have unequal couplings to $u_L$ and $d_L$, the tension between the UA2 dijet constraint and the CDF excess is alleviated.  Flipping the sign of the coupling to either $u_L$ or $d_L$ (but not both) has no affect on the resonant production cross section, but does affect the $WZ'$ production cross section at the Tevatron, because there are two interfering tree-level diagrams for a given partonic initial state, one with an exchanged $u$-quark in $t$-channel and one with an exchanged $d$-quark.  Holding the magnitudes of the $Z'$ couplings to quarks fixed at the value that saturates the UA2 bound, the $WZ'$ cross section at the Tevatron jumps from 1.4 pb to  5.2 pb with a sign flip, more than large enough to explain the $l\nu jj$ excess.  We get a similar result when we perform the analogous exercise for a $Z'$ whose couplings are proportional to those of the $Z$ in the SM (in which case the couplings to $u_L$ and $d_L$ do have opposite sign).  If the $Z'$ couples {\em only} to $d_L$ the exercise gives an even larger allowed $WZ'$ cross section, 10.8 pb.
However, for the $Z'$ to have non-identical couplings to the left-handed up and down quarks,  it must couple to the sector of electroweak symmetry breaking sector in some non-trivial way, as we have discussed in section \ref{sec:nabiv}.  In the context of a concrete model, this might lead to additional diagrams in the $WZ'$ production amplitude, and would certainly modify predictions for precision electroweak observables, both issues that would require thorough investigation.

It seems that, in a well motivated $Z'$ model, it is hard to achieve the full cross section observed by CDF, while maintaining consistency with dijet bounds from UA2.  However, if the efficiencies for the $Z'$ are different than those for the scalar resonance considered by CDF the required cross section at CDF may be smaller.  In addition there may have been statistical fluctuations at both experiments.  We now point out other channels that can be used to test both the $Z'$ model, and \emph{any} model that attempts to explain the $Wjj$ excess.   We urge that the necessary analyses be carried out at the Tevatron.

\subsubsection{Three approximate analyses - motivation for more precision}

There are several complementary analyses that have also been carried out by CDF: the search for diboson events in the $jj+\met$ channel~\cite{Aaltonen:2009fd}, and an analysis of $\gamma+$jets~\cite{cdfphotonjets}.  We discuss the constraints each may apply on the $Wjj$ signal below.  We find some tension between the $Wjj$ excess and both the $jj+\met$ and the $\gamma+$jets analyses, but we emphasize that these analyses were not designed for the $Z'$ signal and so the efficiency for each of these analyses to pick up an effective $Z'$ is unknown.  We will attempt to estimate the efficiencies from existing results, but urge dedicated analyses be carried out.  These cross checks in other channels are non-trivial tests of a broad class of explanations of the $Wjj$ excess, not just the effective $Z'$.

\subsubsection*{1: Jets + $\met$}

The first analysis ($jj+\met$) is based on $\mathcal{L}_{\met} = 3.5\ \mathrm{fb}^{-1}$ and has selection cuts of $\met > 60\ \gev$, exactly 2 jets with $E_T>25\ \gev$ and $|\eta|<2.0$, there is \emph{no} lepton requirement or veto.  These cuts are to be compared to those of the $Wjj$ analysis which also requires exactly 2 jets, with $E_T>30\ \gev$ and $|\eta|<2.4$, $\met > 25\ \gev$ and exactly 1 isolated lepton with $p_T> 20\ \gev$.  The lack of a lepton veto in the $jj+\met$ analysis means it has the potential to observe some of the excess events seen by the $Wjj$ analysis.  Furthermore, in many instances - including $Z'$s - the $Wjj$ signal will be accompanied by a $Zjj$ signal which will fall into the $jj+\met$ analysis.  The relative efficiencies of the two analyses to any new physics explanation of the $Wjj$ excess must be determined by a full Monte Carlo study, but here we attempt to estimate it and show that in many cases it one would expect a substantial number of events in $jj+\met$.

The number of events the $jj+\met$ analysis would see is
\begin{equation}
N_{\met} = {\mathcal L}_{\met}\times  (\sigma^{WZ'} \epsilon_{\met}^{WZ'}+\sigma^{ZZ'} \epsilon_{\met}^{ZZ'})~.
\end{equation}
In order to estimate the efficiencies $\epsilon_{\met}^{WZ'}$ and $\epsilon_{\met}^{ZZ'}$, we assume that the efficiency is not a rapidly changing function of $m_{jj}$, which appears reasonable based on the backgrounds seen in the $jj+\met$ analysis.  Thus, we make the working assumption that 
$\epsilon_{\met}^{WZ'}$  and $\epsilon_{\met}^{ZZ'}$ differ from the efficiencies for SM diboson events only due to branching ratio differences.  That is, we take
\begin{eqnarray*}
 \epsilon_{\met}^{ZZ}& \simeq  & 2\times BR(Z \rightarrow q {\overline q}) \times \epsilon_{\met}^{ZZ'}\\
 \epsilon_{\met}^{WW} & \simeq  & 2 \times BR(W \rightarrow q' {\overline q}) \times \epsilon_{\met}^{WZ'}~.
\end{eqnarray*}
Using  $\epsilon_{\met}^{ZZ}=2.9 \times 10^{-2}$ and $\epsilon_{\met}^{WW}=2.5 \times 10^{-2}$, the Monte Carlo values reported in \cite{Aaltonen:2009fd}, we find $\epsilon_{\met}^{ZZ'}\approx0.021$ and $\epsilon_{\met}^{WZ'}\approx0.018$.  Thus, the excess seen in $Wjj$ should predict approximately 250 events in $jj+\met$, and if the model also predicts a non-zero $Zjj$ rate there will be additional events.  For our $Z'$ explanation we use MadGraph to find that $\sigma^{ZZ'}$ is smaller than $\sigma^{WZ'}$ by approximately a factor of 3 for a 150 GeV $Z'$ and so we estimate the additional number of events from $ZZ'$ production to be approximately 100, for a total of 350 events.  

There are many caveats associated with this ``analysis" and we present it merely as motivation for the analysis to be done.  For instance, the $jj+\met$ analysis only studies the $m_{jj}$ distribution up to $m_{jj}=160\ \gev$ and although the Gaussian peak seen in the $Wjj$ study is centered at $150\ \gev$ there is evidence~\cite{cdfthesis} that the underlying mass scale may be higher.  If the systematic shift in reconstructed mass is different in the $jj+\met$ analysis these extra events may be beyond the reach of the present analysis.  If the new physics explaining $Wjj$ has kinematics such that the neutrino from the $W$ is always soft then these events will not pass the $\met$ cut of the $jj+\met$ analysis, or will do so with lower efficiency than our estimate.  

\subsubsection*{2: Jets + $\met$}

We present another estimate of the cross-talk between the two analyses that suffers from different approximations, and caveats.  Rather than assume the efficiencies at $m_{jj}=150\ \gev$ can be estimated from the diboson we signal we estimate it by comparing backgrounds in the two experiments.  In taking this approach we have to assume the two analyses would both reconstruct the mass of the dijet system to be $\approx 150\ \gev$.

The combined fit of reference \cite{Aaltonen:2011mk} gives the total number of excess $\ell\nu j j$ events as $253 \pm 42 \pm 38$, where the separate uncertainties are for the numbers of electron and muon events. We take $N_{\ell} = 253$, with the understanding that it comes with a roughly 25\% uncertainty.  
The ratio of $Z'$ events expected to show up in the two analyses is
\begin{equation}
\frac{N_{\met}}{N_{\ell}} = \left( \frac{{\mathcal L}_{\met}}{{\mathcal L}_\ell}  \right) \left( \frac{\sigma^{WZ'} \epsilon_{\met}^{WZ'}+\sigma^{ZZ'} \epsilon_{\met}^{ZZ'}}{\sigma^{WZ'} \epsilon_\ell^{WZ'}+\sigma^{ZZ'} \epsilon_\ell^{ZZ'}} \right)~.
\end{equation}
Because of a cut on $\met$ and a veto on dileptons reconstructing a $Z$,  $ZZ'$ events will rarely be selected by the $jj\ell+\met$ analysis.  We neglect whatever small contribution  arises from $Z\rightarrow \tau \tau$ events and approximate $\epsilon_{\ell}^{ZZ'}$ to be zero.  We expect that the ratio of the numbers of $WZ'$ events that pass the $jj+\met$ and $jj\ell+\met$ cuts is well approximated by the ratio of $N^{W+jets}_{\met}$ and $N^{W+jets}_\ell$, where these are the numbers of $W+$jets background events selected by the two analyses, with dijet mass around $m_{jj} = 150$ GeV.  That is, we assume
\begin{equation}
\left( \frac{{\mathcal L}_{\met}}{{\mathcal L}_\ell}  \right) \left( \frac{\sigma^{WZ'} \epsilon_{\met}^{WZ'}}{\sigma^{WZ'} \epsilon_\ell^{WZ'}} \right) \simeq \frac{N^{W+jets}_{\met}}{N^{W+jets}_\ell}~.
\end{equation}
Of course, a full Monte Carlo study would be required to confirm that this is a good approximation.  

Using Figure 1 of Ref.~\cite{Aaltonen:2011mk}, and the supplementary material to Ref.\cite{Aaltonen:2009fd} available at \cite{cdfwebmet}, we sum the event counts between $m_{jj} = 136$ GeV and $m_{jj} = 160$ GeV to estimate $N^{W+jets}_\ell = 690$ and $N^{W+jets}_{\met} = 1340$.  Temporarily setting $\sigma^{ZZ'}=0$, we can see already that this exercise leads us to expect around 490 events to show up in the $jj+\met$ analysis based on $WZ'$ events alone.  For any  $WX$ explanation of the $\ell\nu j j$ excess, where $X$ decays to dijets, we should expect about this many events even in the absence of $ZX$ production, with the following caveats: the estimate assumes that the relative efficiencies for $W+$jets background events are  similar to that for $WX$ events, and that the $jj$ resonance is reconstructed to the same mass in each analysis.  Even if these assumptions are correct the estimate comes with an uncertainty of at least $\sim 30$\%. 

Adding in the additional events due to the non-zero $ZZ'$ cross section requires us to make the same assumption as before, that the efficiency for $ZZ'$ can be extracted from  that for $ZZ$. Doing so we estimate an additional 180 events, giving a total of 670 events in $jj+\met$.

Are these numbers consistent with the findings of Ref.~\cite{Aaltonen:2009fd}? 
Using the more conservative results from the first analysis,
if we take $\sigma^{WZ'}$ to be 1.4 pb instead of 4 pb, for consistency with the UA2 bound, the $\sim 350$ total events become $\sim 120$ total events.
The dijet mass resolution is such that these events would be spread out over several bins in  Figure 2 of Ref. \cite{Aaltonen:2009fd}, each of which contains on the order of $\sim 1000$ events.  So, it seems that this number of $Z'$-induced events is reasonably consistent with the data.  
It seems less plausible that $\sim 350$ events could escape notice.  For example, taking the dijet resolution to be 15 GeV and centering a Gaussian peak at 150 GeV, one expects 60, 73, and 68 extra events in the last three bins of Figure 2 of Ref. \cite{Aaltonen:2009fd}, where no excess is seen
(if the peak is at 160 GeV the numbers  become 31, 54, and 71, and if the peak is at 170 GeV they are 11, 26, and 48).  
The statistical uncertainty in each bin is about 30 events, which is larger than the systematic uncertainty associated with the electroweak background estimation for all but the last bin, where the systematic uncertainty is about 40 events. So, it seems that there is tension, and because most of the events come from $WZ'$ production rather than $ZZ'$ production, there appears to be tension for any  $WX$ explanation of the $l\nu j j$ excess with $ \sigma^{WX} = 4$ pb, unless the efficiency for $WX$ events is significantly smaller than we have estimated.  However, a dedicated analysis by the collaboration would be required to say something more definite. 

\subsubsection*{3: Jets + $\gamma$}

If the $jj$ resonance in $Wjj$ is made in association with a $W$ then it can also be made in association with a $\gamma$.  This is not true if the $Wjj$ is itself resonantly produced~\cite{Eichten:2011sh}.  This brings us to the final associated channel~\footnote{We thank Matt Reece for bringing this to our attention.}: the CDF~\cite{cdfphotonjets} $\gamma+$jets analysis, which is based upon 4.8 fb$^{-1}$.  This analysis selects events with a central, isolated photon with $E_T>30\ \gev$, and one or more jets with $E_T>15\ \gev$.  For the case of two or more jets the invariant mass of the two leading jets is studied and no discrepancy from the SM is observed.

For a $Z'$ of 150 GeV with $\sigma_{WZ'} = 4$ pb at the Tevatron, 
we find that the $\gamma Z'$ production cross section, after cuts, is 1.5 pb, to be compared with the rates in the SM of $\sigma_{\gamma Z}\approx 1.8$ pb and $\sigma_{\gamma W}\approx 1.2$ pb.  Since we do not know the efficiency of this analysis to the $\gamma Z'$ signal we assume  $100\%$ efficiency of the signal, to be conservative (and alternatively an efficiency of $68\%$, based on the search for new physics in the exclusive $\gamma+\met$ channel, which has a slightly higher photon $E_T$ requirement~\cite{gammamet}).  Thus, we expect  7200 ($\sim 4900$) total events, distributed over several bins centered around 150 GeV.    Assuming a dijet mass resolution of $\sim 14\ \gev$~\cite{Aaltonen:2011mk} this predicts 1900 (1300) events in 10 GeV wide bins on either side of 150 GeV.  

We compare the predicted number of events in the 150-160 GeV bin to the number observed~\cite{privatecommunication}, which is $\sim 10^4$.  The data agrees well with the SM prediction and in the bin in question there is $\approx 5\%$ combined systematic and statistical uncertainty in the ratio (data-background)/data~\cite{cdfphotonjets}.  Thus, for a (likely overly optimistic) $100\%$ efficient analysis there appears to be tension between the $\sim 19\%$ correction and the uncertainty in the prediction.  However, for the more realistic $\sim 70\%$ efficient analysis it is possible that a $\sim 13\%$ correction would have gone unnoticed.  These numbers all assume that the necessary cross section to explain the $Wjj$ excess is 4 pb.  If instead this is an overestimate due to upward fluctuations in the data, or increased efficiency for the $WZ'$ signal, the corrections above would be correspondingly reduced and thus could more easily have been missed.  Furthermore, if the UA2 constraint is satisfied the $Wjj$ cross section at the Tevatron must be $\ltap1.4$ pb and the corrections from $\gamma Z'$ become small enough to avoid detection. Alternatively, if the dijet mass resolution is larger the signal will be spread over more bins and could be missed. 

It seems that a dedicated search for a feature in the dijet spectrum in association with a hard photon has potential to provide non-trivial constraints on scenarios that explain the $Wjj$ excess through associated production of a $W$ boson and a $jj$ resonance.  Another study~\cite{Jung:2011ua}, reaches the conclusion that the $\gamma jj$ channel provides a more stringent constraint on  $Z'$ explanations of the $Wjj$ excess than the UA2 dijet constraint.

\subsection{$A_{FB}^{t}$ signal}

In this section we explore whether an effective $Z'$ can explain the recent anomaly in the top quark forward-backward asymmetry, whilst remaining compatible with other observations of top quark properties at the Tevatron and LHC.

Using 5.3 fb$^{-1}$ of data the CDF collaboration has measured the top quark forward-backward asymmetry (in the $t\bar{t}$ frame) to be $\Afbtt = 0.158\pm0.075$ where the error is a combination of statistics and systematics~\cite{Aaltonen:2011kc}.  This is consistent with, although higher than, the SM prediction  $0.058\pm 0.009$, dominated by NLO QCD.  However, the asymmetry is a function of $t\bar{t}$ invariant mass and is very different at high and low $M_{t\bar{t}}$,
\bea
\Afbtt(M_{t\bar{t}}<450\ \gev)&=&-0.116\pm 0.153 \nonumber \\
\Afbtt(M_{t\bar{t}}\ge450\ \gev)&=&0.475\pm 0.114~.
\eea
The SM prediction is $0.040\pm 0.006$ and $0.088\pm0.013$ in the same regions.  At high invariant mass the observed asymmetry is $3.4\,\sigma$ higher than the SM prediction.

At the same time both D0 and CDF have made measurements of the $t\bar{t}$ production cross section.  The most precise determination from D0~\cite{Abazov:2011mi} uses 5.3 fb$^{-1}$ of data and finds $\sigma_{t\bar{t}}=7.78^{+0.77}_{-0.64}$ pb which is consistent with the most precise CDF determination~\cite{cdfttxsec} of $\sigma_{t\bar{t}}=7.50\pm0.48$ pb which uses 4.6 fb$^{-1}$.  These are both in good agreement with the SM prediction of $7.46^{+0.66}_{-0.80}$ pb~\cite{Langenfeld:2009wd}, where in all cases a top quark mass of 172.5 GeV was assumed.  Furthermore, the recent observation of single-top production at the LHC~\cite{cmssingletop}, and its observation at the Tevatron~\cite{Group:2009qk}, are both in agreement with the SM predictions.  There are also no significant excesses at either the Tevatron~\cite{Aaltonen:2008hx,Aaltonen:2009nr} or the LHC~\cite{lhcdilepton} in like-sign dilepton events, a channel in which $tt$ production would show up.

Thus, the challenge to any new physics explanation of the discrepancy in $\Afbtt$ is to explain it without also predicting large corrections to other top quark properties.  Furthermore, in many models there are associated signals in non-top channels.  This is also the case for an effective $Z'$, which contributes to the dijet rate.  

In a single vector-quark model with $M_{Z'} = 150$ GeV, we find that the couplings required to explain $\Afbtt$ lead to serious tension with the measured single-top production cross section and with limits  on same-sign top production, and it is possible that the model is ruled out.  In a more general model with the same $M_{Z'}$, the tension from single-top production can be ameliorated, but one is left with a rate for same-sign top production at the LHC that seems inconsistent with data.  It may be possible to sidestep this constraint in a more complex model with two $Z'$ gauge bosons.

For a heavier $Z'$, the same-sign top problem persists provided that $\Afbtt$ is being explained by t-channel $Z$ exchange.  In the single-vector quark model, flavor-diagonal $Z'$ couplings are inevitable, and for a heavy $Z'$ with $M_{Z'} > 2 M_{top}$, the s-channel $Z'$ exchange dominates over t-channel $Z'$ exchange.  It is  possible to explain $\Afbtt$ with a s-channel exchange of a heavy $Z'$, but the resonant $t{\bar t}$ production tends to lead to dramatic distortion of the $t{\bar t}$ invariant mass spectrum.  A way out is to imagine that the $Z'$ is extremely wide due to decays to non-SM particles, but we find that this approach is not entirely successful.

\subsubsection{UV model}
From the effective theory perspective, we will introduce a coupling of a $Z'$ to right handed up- and top- quarks. However, unlike previous examples, here, to have sizable couplings, we are expected to be integrating out heavy fields with masses comparable to the interacting states in the theory (namely, the top quark). Thus, we should include from the outset the full UV theory, as additional corrections involving the top quark can be important (such as corrections to SM couplings from operators such as (\ref{eq:shiftzwcoupling})).

The simplest  UV realization for an effective $Z'$ for the $\Afbtt$ anomaly introduces a single heavy vector-like quark $(T, T^c)$, where $T^c$ has $U(1)'$ charge $-1$ and SM quantum numbers of the RH up quarks $u_i^c$, and $T$ is its vector parter. We will also discuss generalizations of this minimal model.  By giving $T$ flavor-violating $\phi$-kawa's we get flavor violating $Z'$ couplings.    Considering only $\phi$-kawa's involving the up and top quarks and ignoring the up quark Yukawa, the relevant Lagrangian terms are
\be
\mathcal{L} \supset - \left( \tilde{y}_t t^c t H +\mu T T^c +\lambda_u u^c T \phi +\lambda_t t^c T \phi \right)~.
\label{eq:tfblagrangian}
\ee
After both the scalars get vevs, $\langle H\rangle = v$ and $\langle \phi\rangle = w$, the fermions mix.  By rotating the $u$ and the $u^c$ quarks separately we can diagonalize the mass matrix, i.e. $\vec{u}=V \vec{u}'$ and $\vec{u^c}=V^c \vec{u^c}'$. Because of the large mass of the top quark we cannot neglect the left-handed mixing $V$ as we did for the application to $B_{d,s}$ mixing.    The mass eigenstates have masses 0 and $\left[\frac{1}{2} \left(\Delta^2 + \tilde{m}_t^2 \pm \sqrt{(\Delta^2-\tilde{m}_t^2)^2 + 4 |\lambda_t|^2 w^2 \tilde{m}_t^2}\right)\right]^{1/2}$ where we define $\Delta^2 = (|\lambda_u|^2+|\lambda_t|^2)w^2 +\mu^2$, and $\tilde{m}_t=\tilde{y}_t v$.  Due to mixing and the requirement that there is a top-like state with mass $\approx 170$~GeV, the top quark Yukawa, $\tilde{y}_t$, will be different from its SM value. 

From the rotation matrices it is possible to determine the effective $Z'$ couplings to SM fermions.  Using the notation ${\mathcal L} \supset (g_{ij}^L\; {\bar u}_i {\bar \sigma}^\mu u_j  - g_{ij}\; {\bar u^c}_i {\bar \sigma}^\mu u^c_j) {Z'}_\mu$, the non-zero couplings are
\bea
g_{uu} & = & g' |V^c_{31}|^2  \nonumber\\
g_{ut} = {g_{tu}}^* &  =  &  g' {V^c_{31}}^* V^c_{32} \nonumber\\
g_{tt}  & = & g' |V^c_{32}|^2 \nonumber\\
g_{tt}^{L} & = & g' |V_{32}|^2 ~.
\label{eq:zpcouplings}
\eea
In the minimal case of a single $T$, unitarity relates the couplings, $g_{uu}g_{tt}= |g_{ut}|^2$.  This relation will be altered if there are multiple vector-like quarks that mix with the SM fields, in which case the various couplings become independent parameters.  We do not discuss this more general model in detail, but below we illustrate where our results would be altered.

\subsubsection{Calculation of {\Afbtt}}

Exchange of the $Z'$, both in the t- and s- channels, contributes to $t\bar{t}$ production and to the forward-backward asymmetry.  After the inclusion of the $Z'$ the differential cross section is ~\cite{Cheung:2009ch,Cao:2010zb},
\bea
\frac{d\sigma}{d\cos\theta} & = & \frac{\pi \alpha_s^2\,\beta}{9\hat{s}}\left(2-\beta^2+\beta^2\cos^2\theta\right) \nonumber \\
&+& 
\frac{\beta}{18\hat{s}^2}\frac{\alpha_s |g_{ut}|^2}{t-\mzp^2}\left(2(u-m_t^2)^2+2\hat{s}m_t^2 +\frac{m_t^2}{\mzp^2} ((t-m_t^2)^2+\hat{s}m_t^2) \right) \nonumber \\
&+&
\frac{\beta}{128\pi \hat{s}}\frac{|g_{ut}|^4 }{(t-\mzp^2)^2}\left(4 (u-m_t^2)^2+\frac{m_t^4}{\mzp^4}(4\hat{s}\mzp^2+(t-m_t^2)^2) \right) \nonumber \\
&+&
\frac{\beta  (g_{uu}g_{tt})^2}{32\pi \hat{s} }\frac{(u-m_t^2)^2}{(\hat{s}-\mzp^2)^2+\mzp^2\Gamma_{Z'}^2}\nonumber \\
&+& \frac{\beta  g_{uu}g_{tt}  |g_{ut}|^2}{96\pi \hat{s}}\frac{(\hat{s}-\mzp^2)(\frac{m_t^4}{\mzp^2}\hat{s}+2(u-m_t^2)^2)}{(t-\mzp^2)((\hat{s}-\mzp^2)^2+\mzp^2\Gamma_{Z'}^2)}~\\
&+& \frac{\beta  (g_{uu} g_{tt}^L )^2}{32\pi \hat{s}}\frac{(t-m_t^2)^2}{(\hat{s}-\mzp^2)^2+\mzp^2\Gamma_{Z'}^2}\nonumber~\\
&+& \frac{\beta  g_{uu} g_{tt}^L  |g_{ut}|^2}{96\pi\hat{s}}
\frac{
m_t^2 (\hat{s} - \mzp^2)(
2 \hat{s}  +  (t-m_t^2)^2/\mzp^2 
)}
{(t-\mzp^2)((\hat{s}-\mzp^2)^2+\mzp^2\Gamma_{Z'}^2)}
\nonumber~\\
&+& \frac{\beta   g_{uu}^2 g_{tt}^L g_{tt} }{16 \pi{\hat s}}\frac{m_t^2 \hat{s}}{(\hat{s}-\mzp^2)^2+\mzp^2\Gamma_{Z'}^2}\nonumber~.
\label{eq:diffxsec}
\eea
Where $\beta=\sqrt{1-4m_t^2/\hat{s}}$,  $t=m_t^2-\frac{\hat{s}}{2}(1-\beta\cos\theta)$, and $u=m_t^2-\frac{\hat{s}}{2}(1+\beta\cos\theta)$. 
The first term is LO QCD, the third is $Z'$ exchange in the t-channel, and the fourth, sixth and eighth are $Z'$ exchange in the s-channel, where the latter two terms account for the fact that the $Z'$ may have non-negligible coupling to the left-chiral top quark.  The other terms come from interference,  between QCD and t-channel $Z'$ exchange for the second term, and between $Z'$ exchange in the s- and t- channels for the fifth and seventh terms.  
In the minimal model  the unitarity relationship (\ref{eq:zpcouplings}) between $g_{uu}, g_{tt}$ and $g_{ut}$ allows us to express the first five terms in the cross section in terms of  $|g_{ut}|$, but if more than one vector quark mixes, $g_{uu}, g_{tt}$ and $g_{ut}$  are independent. 
With this additional freedom we can smoothly transition between the t-channel exchange of a flavor-violating $Z'$~\cite{Jung:2009jz} and  the s-channel exchange of a flavor-conserving $Z'$ ~\cite{Cao:2010zb}. 

The non-negligible mixing among the left-handed quarks means that $\phi$ will inherit a flavor-violating $u-t$ coupling as well.  For simplicity we will neglect the effects of $\phi$ in our discussion below.  We expect our quantitative results to be valid provided the $\phi$ particle is somewhat heavier than the $Z'$, potentially requiring a large quartic coupling.  

\begin{figure}[t] 
   \centering
   \includegraphics[width=0.6\columnwidth]{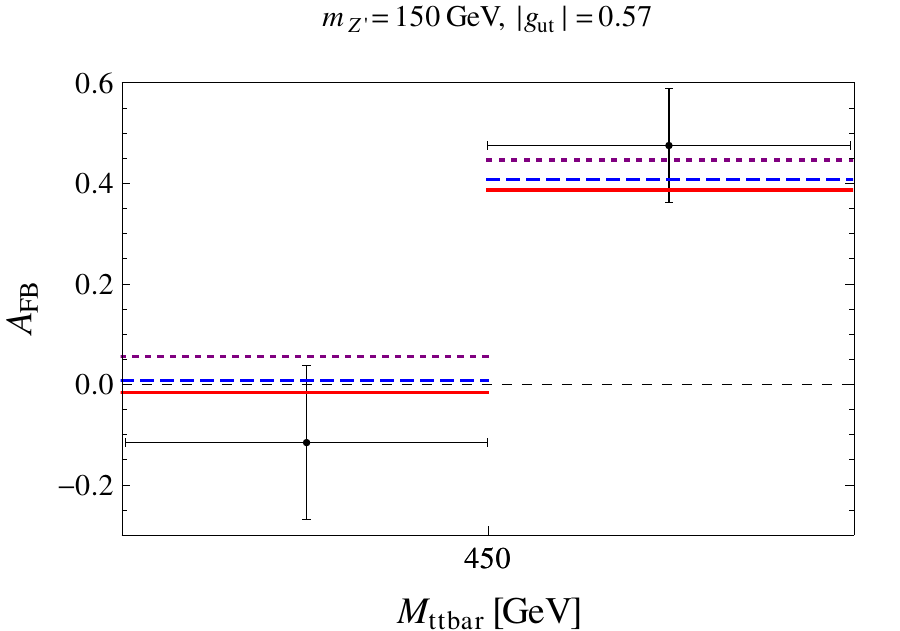} 
   \caption{The predicted top forward-backward asymmetry in the  $t{\bar t}$ rest frame, for $M_{t\bar{t}}<450$ GeV and $M_{t\bar{t}}>450$ GeV.  
 The dotted purple line has all couplings but $g_{ut}$ turned off, the dashed green line is with the minimal-model constraint $g_{uu}g_{tt}= |g_{ut}|^2$ and $g_{tt}^L$ turned off, and the solid red line is with the minimal-model constraint and $g_{tt}^L$ turned on. The contribution to the asymmetry from NLO QCD is not included.}
   \label{fig:mttafb}
\end{figure}

The final three terms in Equation~(\ref{eq:diffxsec}) depend on  the mixing among the left-handed quarks.  
As a first step in our study of $\Afbtt$ we neglect this mixing and set $g_{tt}^L$ to zero.    In Figure~\ref{fig:mttafb} we show that a relatively light $Z'$,  $\mzp=\ 150\;\gev$, fits the forward-background asymmetry well for $|g_{ut}| = 0.57$.
%%%%%%%%%
%%%%%%%%%
In this figure we show the result when the minimal model constraint $g_{uu} g_{tt} = |g_{ut}|^2$ holds, along with the result when $g_{uu}$ is set to zero, so that only $t$-channel $Z'$ exchange contributes.

A caveat is that the parton-level asymmetry reported by CDF uses an unfolding procedure with built-in model assumptions.  To more rigorously determine how well the $Z'$ model can explain the data, one would need to perform a full Monte-Carlo simulation including parton showering and hadronization,  top reconstruction, and detector effects, and then compare to CDF reconstructed data.  This procedure has been carried out for a variety of models in \cite{Gresham:2011pa}.

\subsubsection{UV couplings for $\Afbtt$}

Can a  $|g_{ut}|$ of the size preferred by $\Afbtt$   be generated by mixing with a single vector-quark? In a search for viable model parameters, the following constraints come into play.
For a 150 GeV $Z'$ that couples to right-handed up quarks, the constraint on resonant dijet production from UA2 translates to $g_{uu} \lsim 0.24$.  In the minimal model this immediately tells us that $g_{tt}$ must be rather large, $g_{tt} \gsim 1.35$ for $g_{ut} = 0.57$.  From Equation~(\ref{eq:zpcouplings}), we see that even for maximal $t^c-T^c$ mixing, $|V^c_{32}|^2 \sim 1/2$, we need a large $U(1)'$ gauge coupling, $g' \gsim 2.7$.  

The coupling of the top quark to $Wb$ will be reduced by left-handed $t-T$ mixing.  The value of $|V_{tb}|$ is constrained by single-top production (whereas  top decay branching ratios only provide a constraint once CKM unitarity is assumed).  Other constraints come from the $B^0_d$ and $B^0_s$ systems, in which the SM contributions to the mass splittings are proportional to $|V_{tb}|^2$.  A rigorous estimation of the allowed range of $|V_{tb}|$ would require a global fit with the CKM unitarity assumption relaxed, including the effects of the heavy $T$ quark (and any other extra quarks,{\em e.g.} the right-handed down quarks may also mix as in Section~\ref{subsec:dimuons}), and this is beyond the scope of this work.  We simply require $|V_{tb}| > 0.8$, which is roughly consistent with the combined CDF and D0 constraint from single top production, $|V_{tb}| = 0.88 \pm 0.07$ \cite{Group:2009qk}.   
As discussed below, the $Z'$ gives new contributions to single-top production. So, somewhat smaller values of $|V_{tb}|$ may in fact be consistent with the data once $Z'$-induced single-top production is included, but at some point one expects tension with $\Delta M_{B_d}$ and $\Delta M_{B_s}$.  The $|V_{tb}|$ constraint requires couplings to be even larger than demanded by UA2.  

The $T$ quark must also be heavy enough to have evaded detection.  Through the left-handed $t-T$ mixing, the heavy quark can decay to $ bW$, and if it is heavy enough, $tZ$.  It can also decay to $Z' u$ and $Z'^{(*)} t^{(*)}$, with the $Z'$ subsequently decaying to $u {\bar u}$. If the $bW$ branching ratio were 100\%, a CDF analysis \cite{cdftprime} would constrain the $T$ mass to be above 335 GeV at 95\% CL.  However, depending on the parameters of the model, the $T$ quark may well prefer to go to $Z' u$, leading to all-hadronic final states.   Events with $T{\bar T} \rightarrow t{\bar t} Z' Z'$, with each $Z'$ subsequently decaying to two jets, might show up as an excess in the $t{\bar t}$ + jets,  but with four extra hard jets the top reconstruction  efficiency would likely  be degraded.  We require $m_T > 300$ GeV, but strictly speaking, for any given parameter point a detailed analysis would be required to determine whether the corresponding mass is consistent with data, and it may be that lighter $T$'s are viable.  

A sample point that satisfies the dijet, $V_{tb}$, and $m_T$ constraints  has  $g' = 2.88$, $\lambda_u = 1.87$, $\lambda_t = 3.68$, $\mu = 236$ GeV, and ${\tilde m}_t = 218$ GeV, giving $|V_{tb}| = 0.86$, $g_{uu} = 0.23$, $g_{ut} = 0.57$, and $m_T = 310$ GeV.  For this parameter point,  the heavy quark couples to $Z't (-1.27)$, $Z't^c(-1.32)$, $Z' u^c(-0.53)$, $Wb(-0.24)$, and  $Zt(0.16)$, where coupling strengths are given in parenthesis.  The heavy quark also has relatively large couplings to $\phi u$ and $\phi t$, which may be relevant if $\phi$ is sufficiently light.  
For this parameter point the left-handed $t-T$ mixing is not small, and we find $g_{tt}^L = 0.77$.  In Figure \ref{fig:mttafb} we see that  including this coupling  does not significantly affect the top forward-backward asymmetry.  

\subsubsection{$\sigma_{t {\bar t}}$, same-sign tops, and single top production}

As well as explaining the observed discrepancy in $A_{FB}$ the $Z'$ must not introduce large deviations from the SM in other channels.  To check this, we implement the model in MadGraph/MadEvent 4.4.56 \cite{Alwall:2007st} (using CTEQ6L1), taking $m_t=172.5\ \gev$.  
We first investigate the effect of the $Z'$ on the $t\bar{t}$ cross section.  For the SM we find the cross section at the Tevatron to be 5.93 pb.  Matching to the NLO prediction requires the introduction of a K-factor of $K= 1.26^{+0.11}_{-0.13}$, and we assume this K-factor is the same for NP.   For $M_{Z'} = 150$ GeV and $|g_{ut}| = 0.57$ this leads us to estimate cross sections of $6.92^{+0.61}_{-0.74}$ pb, $6.40^{+0.57}_{-0.69}$ pb, and $6.45^{+0.57}_{-0.69}$ pb respectively, for the cases with only $g_{ut}$ turned on, the constraint $|g_{ut}|^2 = g_{uu} g_{tt}$ applied with $g^L_{tt}$ turned off, and for the particular minimal-model parameter point discussed above, with $g^L_{tt} = 0.77$.    Compared to the case where the $Z'$ is only exchanged in t-channel, we see that the negative interference between the s- and t-channel $Z'$ diagrams reduces the cross section slightly, and there is some tension with the  D0 and CDF measurements.  It is possible that our assumption that the same QCD K-factor applies to the new physics contribution underestimates the true cross section.   

As shown in \cite{Jung:2009jz}, explaining $\Afbtt$ with t-channel $Z$ exchange leads to a leading-order $M_{t{\bar t}}$ distribution that is apparently too low at low invariant mass, and too high at high invariant mass, see Figure~\ref{fig:spectra}.  The excess at high invariant mass may in fact be resolved by a lower efficiency for high-$M_{t {\bar t}}$ events to be reconstructed, as suggested in \cite{Gresham:2011pa}.  There is also once again the fact that the full NLO calculation has not been done for these models of new physics.  

The same $g_{ut}$ coupling will lead to like-sign top production and a low background same-sign dilepton signal.  
The like-sign cross section only depends on $g_{ut}$ and so is independent of variations on the minimal model. We find that at the Tevatron our benchmark parameter point has a like-sign top production cross section of
\bea
\sigma_{tt+\bar{t}\bar{t}}(\mzp = 150\ \gev)&\approx &\left| \frac{g_{ut}}{0.57}\right|^4\times 0.69 \ \mathrm{pb}~. 
%\\
%\sigma_{tt+\bar{t}\bar{t}}(\mzp = 600\ \gev) &\approx & 0.06 \ \mathrm{pb}~. \nonumber
\eea
The cross section at the  LHC is much larger, in part because the LHC is a p-p collider:
\bea
\sigma_{tt+\bar{t}\bar{t}}(\mzp = 150\ \gev) &\approx &  \left| \frac{g_{ut}}{0.57}\right|^4\times 72 \ \mathrm{pb}~, 
% \\
%\sigma_{tt+\bar{t}\bar{t}}(\mzp = 600\ \gev) &\approx & 12 \ \mathrm{pb}~. \nonumber
\eea
Although the rate at the Tevatron may be low enough to have escaped detection, the approximate upper bound is 0.7 pb~\cite{Aaltonen:2008hx,Cao:2010zb}, the LHC rate appears large.  We estimate that the efficiency $\times$ acceptance of the CMS like-sign dilepton search~\cite{lhcdilepton} is about 20\%.  Combining this with the leptonic branching ratios of the $W$, and ignoring $\tau$ for simplicity although it too will contribute to the SS dilepton signal, we estimate that the our benchmarks gives $\sim 20$  events at CMS.  The CMS analysis sees 0 events, with an expected background of $1.2\pm 0.8$.  We note that, although we have assumed the same lepton selection efficiencies, it is not clear if the efficiencies of a SUSY search apply to our model.  There is however clearly tension for such a light $Z'$.

In a more baroque model with {\em two} $Z'$ gauge bosons it may be possible to get around the same-sign top issue.  Suppose each $Z'$ couples to the right-handed up quarks through mixing with (different) vector-quarks, leading to  separate couplings $g_{ut}^A$ and $g_{ut}^B$.  The amplitude for $uu\rightarrow tt$ has two terms proportional to   $(g_{ut}^A)^2$ and $(g_{ut}^B)^2$ , leaving room for cancellation, whereas for $u{\bar u}\rightarrow t{\bar t}$ the terms are proportional to $|g_{ut}^A|^2$ and $|g_{ut}^B|^2$. 

Another top observable that is affected by the flavor violating $Z'$ coupling is the single top production cross section. There are two production mechanisms: true single top production, proceeding through s- and t-channel $Z'$ exchange,  and $tZ'$ production with the $Z'$ subsequently decaying to $u\bar{u}$.  The later only depends on the $g_{ut}$ coupling and so is a unavoidable prediction of any model that explains $\Afbtt$ by t-channel $Z'$ exchange.  In the first channel, the cross section becomes sensitive to the nature of the model details, since the rate scales as $(g_{uu}g_{ut})^2$.  If the $\phi$-kawa's and vector-quark content are such that $g_{tt}\gg g_{uu}$ the single top production cross section in this channel can be suppressed, but $tZ'$ production can never be avoided.  However, $tZ'$ will contain more jets than SM single-top production and it is not clear  what the efficiencies for this mode will be.  For instance the recent LHC search requires exactly two jets~\cite{cmssingletop}, while the Tevatron analyses employ a sophisticated multivariate technique, and the efficiency for our signal is not known.

At the Tevatron we find the irremovable $tZ'$ production cross section for  our benchmark point to be,
\be
\sigma_{tZ'+\bar{t}Z'}(\mzp=150\ \gev) = \left|\frac{g_{ut}}{0.57}\right|^2 \times  2.0\ \mathrm{pb}~.
\ee
The cross section for true single top production is
\be
\sigma_{tj+\bar{t}j}(\mzp=150\ \gev) = \left|\frac{g_{ut}}{0.57}\right|^2 \left(\frac{g_{uu}}{0.23}\right)^2  \times 4.8\ \mathrm{pb}~.
\ee
Thus, there is clear tension with the Tevatron's result of $\sigma_{t+\bar{t}} = 2.76^{+0.58}_{-0.47}$ pb from the $tZ'$ channel alone, if it efficiently passes the complex multivariate analysis of CDF and D0.  The $Z'$ contribution to true single top can be suppressed to a sufficiently small rate by requiring $g_{uu}\ltap 0.1$ (although this is probably not feasible in the minimal model, because it would require  couplings even larger than those of our benchmark point).  Furthermore, the SM prediction for single top is proportional to $|V_{tb}|^2$ which, as we have discussed, will generally be smaller than its SM value.  Suppressing the SM contribution to single top leaves more room for additional contributions from $Z'$ exchange.

At the LHC the single-top cross sections are
\bea
\sigma_{tZ'+\bar{t}Z'}(\mzp=150\ \gev) &=& \left|\frac{g_{ut}}{0.57}\right|^2  \times 90\ \mathrm{pb} %\nonumber \\
%\sigma_{tZ'+\bar{t}Z'}(\mzp=600\ \gev) &=& \left(\frac{g_{ut}}{0.85}\right)^2  3\ \mathrm{pb}~,
\eea
and
\bea
\sigma_{tj+\bar{t}j}(\mzp=150\ \gev) &=& \left|\frac{g_{ut}}{0.57}\right|^2 \left(\frac{g_{uu}}{0.23}\right)^2 \times 49\  \mathrm{pb}
% \nonumber \\
%\sigma_{tj+\bar{t}j}(\mzp=600\ \gev) &=& \left(\frac{g_{ut}}{0.85}\right)^2 \left(\frac{g_{uu}}%{0.85}\right)^2 51\ \mathrm{pb}~.
\eea
The CMS measurement $\sigma_t\approx 84\pm 30$ pb \cite{cmssingletop} and  SM prediction $\approx 60$ pb again have tension with the unavoidable mode, although we again emphasize that there is a jet veto in the CMS analysis that will affect the efficiency for this signal.  The $tj$ mode  is safe from the LHC constraint once it is below the Tevatron bound.

\subsubsection{A heavier $Z'$}

It is also possible to explain $\Afbtt$ with a $Z'$ that is much heavier than we have considered.  For example, the authors of Ref.~\cite{Gresham:2011pa} consider a benchmark point with $M_{Z'} = 400$ GeV and $g_{ut} = 1.75/\sqrt{2}$, and find a good fit to the CDF results.  For this scenario to work it is important that there is no appreciable contribution from s-channel $Z'$ exchange, which means that to realize it we need to go beyond the minimal model with a single vector quark.  For this heavier $Z'$, we find a total cross section $7.54^{+0.67}_{-0.81}$, where we again use the K factor from NLO QCD.  

True single-top production will necessarily be very suppressed for this parameter point, because the $g_{uu}$ coupling must be small enough so that s-channel $Z'$ exchange can be neglected. The top quark can still be produced in association with $Z'$ as before, but the cross sections are much smaller for this larger $M_{Z'}$,   $\sigma_{tZ'} +   \sigma_{{\bar t}Z'} = 0.1$ pb at the Tevatron and 29 pb at the LHC.   However, the  same-sign tops are produced at an even larger rate than for the $M_{Z'} = 150$ GeV benchmark.  We find $\sigma_{tt+\bar{t}\bar{t}} = 1.2$ pb at the Tevatron and 169 pb at the LHC.  If the $Z'$ carries a flavor charge this constraint is avoided, as considered in~\cite{Gresham:2011pa}.

If the $Z'$ is heavier still, it becomes possible to explain $\Afbtt$ with s-channel $Z'$ exchange \cite{Cao:2010zb}.  In this case the main problem is that a feature in the $M_{t{\bar t}}$ would be expected due to resonant $t {\bar t}$ production, and to suppress this feature the $Z'$ must be very wide.  Unlike the case where $\Afbtt$ comes from a heavy $Z'$ exchanged in t-channel, in this case the minimal model can be used to generate the necessary couplings, although to make the $Z'$ sufficiently wide it presumably must have additional decay modes into non-SM states.  

\begin{figure}[t] 
   \centering
   \includegraphics[width=0.6\columnwidth]{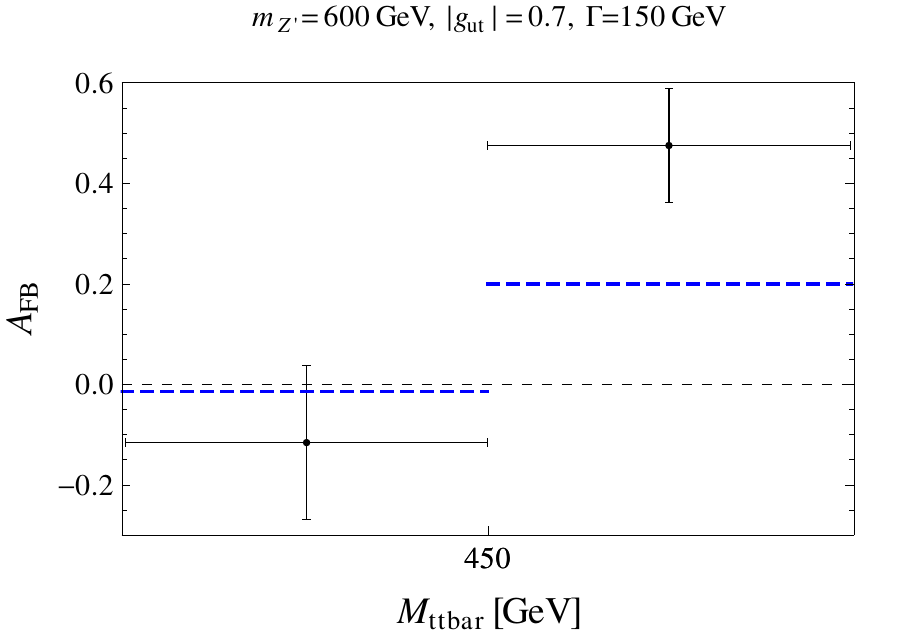} 
   \caption{For the $M_{Z'} = 600$ GeV benchmark, the predicted top forward-backward asymmetry in the  $t{\bar t}$ rest frame, for $M_{t\bar{t}}<450$ GeV and $M_{t\bar{t}}>450$ GeV.  The contribution to the asymmetry from NLO QCD is not included.}
   \label{fig:mttafb600}
\end{figure}

\begin{figure}[t] 
   \centering
      \includegraphics[width=0.48\columnwidth]{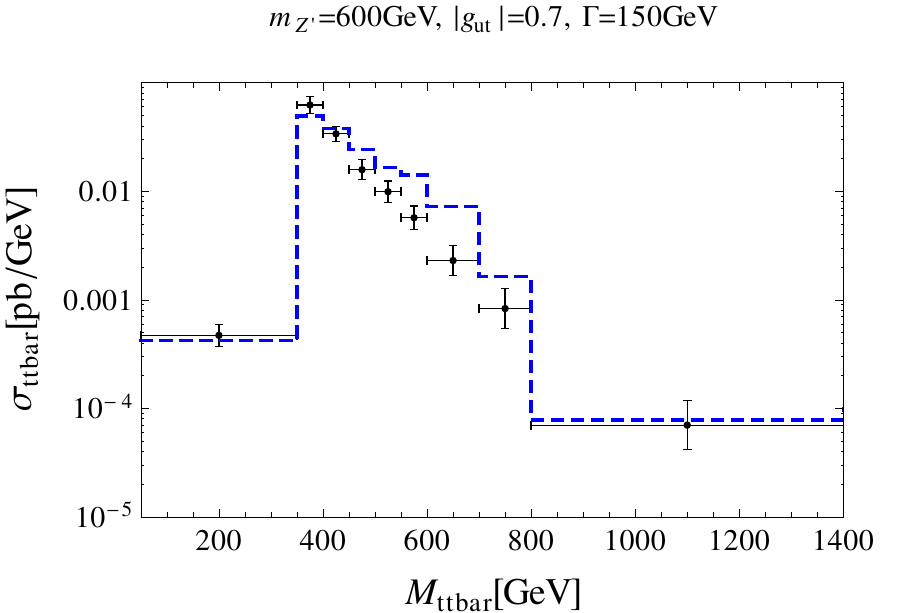} 
   \includegraphics[width=0.48\columnwidth]{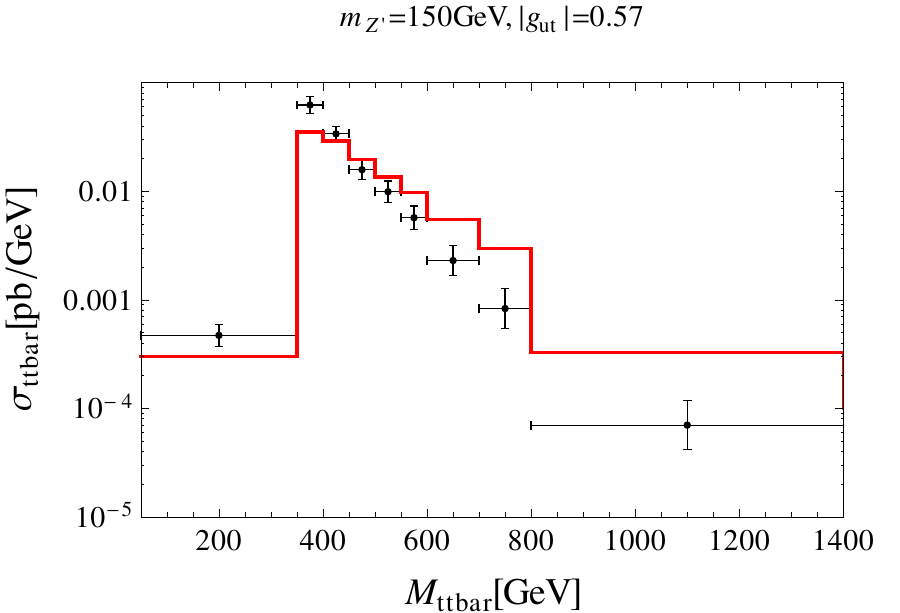} 
   \caption{Predicted $M_{t\bar{t}}$ spectra, with the K factor applied, along with the measured values reported in \cite{Aaltonen:2009iz}. }
  \label{fig:spectra}
\end{figure}

In Figure~\ref{fig:mttafb600} we show the  $\Afbtt$ generated by a 600 GeV $Z'$ with a width $\Gamma_{Z'} = 150$ GeV and a coupling $g_{ut} = 0.7$, with the constraint $|g_{ut}|^2 = g_{uu} g_{tt}$ applied.  We neglect $g_{tt}^L$, which can be much smaller than for the lighter $Z'$, so that the differential cross section is given by the first five terms of Equation~(\ref{eq:diffxsec}).  
%%%%%%%%%
%%%%%%%%%
From the figure, we see that the asymmetry  for this point is lower than that reported by CDF.  If we increase $g_{ut}$  to raise the asymmetry, the $M_{t {\bar t}}$ spectrum becomes more distorted and the total cross section becomes too large.  The $M_{t  {\bar t}}$ distribution for $g_{ut} = 0.7$ is shown in 
Figure~\ref{fig:spectra}.  
%%%%%%%%%
%%%%%%%%%
The excess around $\mzp$ is certainly noticeable, but the distribution may not be much worse than for the $M_{Z'} = 150$ GeV benchmark, which is also shown for comparison.  The caveat from before applies again here:  the efficiencies for these $Z'$ models may not be well represented by those used in determining the $M_{t\bar{t}}$ distribution from data.  

For this point the same-sign tops cross section is only 31 fb at the Tevatron and 5.9 pb at the LHC, so that the tension from same-sign dilepton searches at the LHC is essentially gone. The top-$Z'$ production cross section is $\sigma_{tZ'} +   \sigma_{{\bar t}Z'} = 2$ pb at the LHC, and negligible at the Tevatron.  The cross sections for true single-top production are
\begin{equation}
 \sigma_{tj} +\sigma_{{\bar t}j} =    \left|\frac{g_{ut}}{0.7}\right|^2 \left(\frac{g_{uu}}{0.7}\right)^2 \times 4.0 \;{\rm pb} 
\end{equation}
at the Tevatron and
\begin{equation}
 \sigma_{tj} +\sigma_{{\bar t}j} =    \left|\frac{g_{ut}}{0.7}\right|^2 \left(\frac{g_{uu}}{0.7}\right)^2 \times 30\;{\rm pb} 
\end{equation}
at the LHC.  To be consistent with single-top production at the Tevatron, $g_{uu}$ should be somewhat small, $g_{uu} \ltap 0.2$.

%%%%%%%%%%%%%%%%%%%%%%%%%%%%%%%%
\section{Effective $Z'$s and Dark Matter}\label{sec:dm}
When discussing the prospects for dark matter detection, there is always an underlying assumption that some force will mediate the interaction between dark matter and SM (observable) particles. That can be states present in the SM, such as the Z or Higgs boson, or a nearby cousin, such as the additional scalars in two Higgs doublet models. However, the presence of new forces provides new possibilities.  Scattering can occur via exchange of them, or annihilation into these force carriers can lead to new indirect signals. Thus, the discussion of an effective $Z'$ prompts us to consider the implications for dark matter searches. As we shall see, options open for light WIMP freezeout, modifications of couplings to the SM in direct detection, and new scenarios for indirect detection. 

\subsection{Light Dark Matter and Effective $Z'$s}

Recent experimental results, in particular DAMA and CoGENT have galvanized efforts to understand the possibility of light WIMPs ($m_\chi \lsim 10\ \gev$). One of the central questions for such a light particle is: if it is a thermal relic, how does it maintain equilibrium at such low temperatures? Limits from the invisible width of the Z strongly constrain the existence of light objects with weak charges.

In the presence of an effective $Z'$, however, such concerns are eliminated, if only because the invisible width of such an undiscovered particle is just one of its poorly constrained properties. Dark matter annihilating $\chi \chi \rightarrow Z^{\prime*} \rightarrow SM SM$ can provide a natural freezeout process.

Normalizing to the Lee-Weinberg bound, for WIMPs annihilating through the SM Z requires a mass of $\sim$ 10 GeV. While couplings to the SM must be weaker here, the couplings to WIMPs can be larger, and thus WIMPs lighter than $m_Z/2$ are generally viable here, while in the SM through the Z, they are not.

Alternatively, we can consider  $\chi \chi \rightarrow Z' Z'$. For heavy WIMPs, this can be dialed, simply by adjusting the coupling of the WIMPs. However, for light WIMPs, we have the possibility of freezing out into light, metastable $Z'$s, as in \cite{Finkbeiner:2007kk,Pospelov:2007mp}. Here, however, rather than using the Higgs or kinetic mixing portal, we can have the $Z'$ decay through the effective couplings, providing a new portal for freezing out into dark forces, allowing light WIMPs to have the appropriate relic abundance, naturally.

\subsection{Direct Detection Signatures}
In the presence of an effective $Z'$, there are new possibilities for direct detection mediated by the new force carrier. At the outset, we should hope to estimate what the natural size of cross section would be. We start by assuming the WIMP is a heavy Dirac fermion, and consider the vector-vector interaction, which we can compare with those of the SM.

We assume that the interaction is generated by mixing with a heavy quark of mass $M$ through a $\phi$-kawa $\lambda$, where a vev $v$  breaks the $U(1)$. The natural size of the cross section is
\be
\sigma \approx \frac{16 \pi \alpha_{Z'}^2 \mu_{\chi N}^2 Z_{eff}^2}{\mzp^4} \sin^4 \theta
\ee
where $Z_{eff}$ is the effective charge of the nucleus, which can vary depending on the couplings of the $Z'$ to ordinary matter (for comparison, the usual Higgs-mediated exchange has the standard $Z_{eff}^2 = A^2$, while for SM $Z$ exchange, $Z_{eff}^2 \approx ( (A-Z) - 0.08 Z)^2$). $\sin \theta$ is the mixing angle between the charged and uncharged quark states. $\mu_{\chi N}$ is the reduced mass of the WIMP-nucleus system.

Since we know that $\mzp = g v$ and $\sin \theta \approx \lambda v/M$ we have
\be
\sigma \approx \frac{16 \pi \alpha_{Z'}^2 \mu_{\chi N}^2 Z_{eff}^2}{g^4 v^4} \frac{\lambda^4 v^4}{M^4} = {\lambda^4 \frac{\mu^2_{\chi N}}{M^{4}} Z_{eff}^2}{\pi }.
\ee
Thus we see the important result that {\em for $O(1)$ $\phi$-kawa couplings, the natural scale of the interaction cross section is set by the mass of the heavy quarks, rather than the mass of the $Z'$}. Of course we have assumed that the quarks can be integrated out (so are at least as heavy as the $Z'$), but the $Z'$ could be much heavier. Under the assumption that these fields are weak scale, then the cross section is naturally weak scale as well. This is important, because the standard $\sigma_0$ (cross section per nucleon) for $Z$-exchange is $\sim 10^{-39} {\rm cm^2}$, while current experiments are probing ranges as low as $\sim 10^{-44} {\rm cm^2}$. Thus, even for heavy ($\sim$ TeV) charged quarks, the cross section can be as large or larger than what is expected for Higgs exchange ($\sim 10^{-45} {\rm cm^2}$). Moreover, a light WIMP interacting through an effective $Z'$ would have a much greater chance to be detected in a direct detection experiment, where sensitivity to light particles typically requires cross sections $10^{-42}- 10^{-40}\ {\rm cm^2}$. Moreover, this relationship should hold even for WIMPs with couplings to {\em light} ($\sim$ GeV mass) $Z'$s. This is qualitatively similar to the natural weak-scale cross section found in SUSY models with kinetic mixing for U(1)'s \cite{Cheung:2009qd}.

But beyond the size of these couplings, there are a number of other salient features. Assuming couplings to $u^c$, $d^c$, and $q$ fields, the couplings to protons and neutrons can be expressed in terms of the couplings to quarks, $a_i$, as
\be
\bar p\left[ (2 a_u + a_d)\gamma^\mu\frac{(1 + \gamma_5)}{2} + 3 a_q\gamma^\mu\frac{(1 - \gamma_5)}{2} \right]p \\
\bar n\left[ (a_u +2 a_d)\gamma^\mu\frac{(1 + \gamma_5)}{2} +3  a_q\gamma^\mu\frac{(1 - \gamma_5)}{2}  \right] n .
\ee
This means that proton and neutron couplings $f_p$ and $f_n$ are effectively free parameters. This can have important implications for direct detection experiments, in particular if there is destructive interference. For instance, for $f_n \approx -0.7 f_p$, the light DAMA and CoGeNT regions overlap, while suppressing XENON signals by up to a factor of 100 (but does not address CDMS constraints, which use Si and Ge) \cite{Cheung:2009qd,Feng:2011vu}. In general, the relative sensitivity of light versus heavy targets can be very sensitive to this destructive interference.

A second important point is the possibility of significant spin-dependent interactions. Since the strength of these interactions is only somewhat below $Z$ strength, we can reasonably have spin-dependent interactions which are just below $Z$ strength, as well. However, in for instance SUSY, it is requires significant tuning to generate SD couplings without sizable SI couplings \cite{Cohen:2010gj}. While this has been most studied in SUSY, the reasons are quite general - to avoid over depleting the WIMP in the early universe, an $SU(2)$ charged object must mix with a neutral state. This mixing is achieved by a Higgs coupling of reasonable size, which then mediates a significant SI interaction.

Here, however, the $Z'$, naturally heavier than the $Z$ and with weaker couplings to SM fields, need not have its coupling to the WIMP additionally suppressed, while a small Majorana mass term (presumably from some mixing with a $U(1)'$ neutral state) is adequate to suppress the vector-vector coupling, thus, a sizable SD can arise, and with unspecified proton and neutron couplings $a_p$ and $a_n$.

Finally, we note that if the dark matter {\em does} have $O(1)$ couplings to $\phi$, then it can interact with matter. Specifically, the quark masses depend on $\phi$ as
\be
m_q = m_q^0 \left(1 - \frac{2 v_\phi \lambda^2 \phi}{v_\phi^2 \lambda^2 + \mu^2}\right) \approx  m_q^0 \left(1 - \frac{2 \sqrt{2} g_{eff} \phi}{ M_{Z'}}\right).
\ee
One compares this with the usual relationship for a Higgs
\be
m_q = m_q^0\left( 1 + \frac{h}{v}\right).
\ee
Of course, we do not expect $\phi$ to couple to all quarks, but we might still expect a cross section
\be
\sigma_\phi \approx  \frac{\alpha'_{eff} M_W^2}{\alpha_W M_{Z'}^2} \frac{m_h^4}{m_\phi^4} \times \sigma_h,
\ee
where $\sigma_h$ is a characteristic Higgs scattering cross section typically $\sim 10^{-45}{\rm cm^2}$ per nucleon. Thus, even if the dark matter is a Majorana fermion, and has no spin-independent scattering mediated by the $Z'$, the detection of an effective $Z'$ would give insight into the $\phi$-mediated spin-independent cross section we might expect at a direct detection experiment.

\subsection{Indirect Detection}
In addition to direct detection experiments, dark matter can be detected through its annihilation products. The annihilation through an effective $Z'$ can give rise to different final states, for instance $\tau^+ \tau^-$, that might occur rarely in conventional annihilation modes. While such models cannot avoid the usual helicity suppression for Majorana WIMPs annihilating into SM fermions \cite{Goldberg:1983nd}, it does create a number of interesting new possibilities.

To begin with, let us consider the case of WIMPs annihilating after solar capture. If the coupling is dominantly to first generation quarks, then the light-flavor hadrons produced in annihilation ($\pi^\pm$) can stop in the solar interior before decaying, lowering the energy of the resultant neutrinos and weakening limits compared with heavy flavor \cite{Cirelli:2005gh}. Similarly, the $Z'$ could be dominantly leptophilic, and if coupling to $\mu$ or $e$, would also produce no interesting limits from solar capture. Such mechanisms to limit the solar capture signals can be important in models where the capture rate is high, such as inelastic dark matter and spin-dependent scattering, both of which can be realized with effective $Z'$s.

Effective $Z'$s can have interesting applications when considering the PAMELA/Fermi cosmic ray anomalies, as well. One now-conventional way to generate lepton-rich annihilation final states from a heavy particle is to have the WIMP annihilate dominantly into light ($\lsim$ GeV) mediators, which decay to SM particles \cite{Finkbeiner:2007kk,Cholis:2008vb}. Given the constraints from $\bar p$ measurements, this is an important tool in many models that explain the positron excess \cite{ArkaniHamed:2008qn,Pospelov:2008jd}. Here, however, we can imagine a light $Z'$ which couples dominantly to leptons simply by the nature of its couplings, realizing a more directly leptophilic model as conceived by \cite{Fox:2008kb}, providing a natural alternative to the kinetic mixing approach to light mediators. Finally, we note that if the coupling is leptophilic via heavy $L$ fields (rather than $e^c$ fields), then significant neutrino signals should be present, as well.

In summary, the presence of an effective $Z'$ makes natural a wide range of WIMP phenomenology, including light WIMPs, generalized proton and neutron SI couplings, SD interactions without sizable SI interactions, and modifications to expectations from indirect signals. While the viability of any particular set of couplings (for instance to explain a given anomaly) must be quantitatively studied in context, it is clear that effective $Z'$s can change the phenomenology of WIMPs considerably.

%%%%%%%%%%%%%%%%%%%%%%%%%%%%%%%%%%%%
\section{Discussion}\label{sec:conclusions}
Supplementing the SM by the addition of new $Z'$s is one of the most natural extensions of the SM that can be imagined. However, charging existing SM fermions under such a $Z'$ changes it to one of the most unwieldy and unattractive extensions. The alternative discussed here - to only effectively charge the SM under the $Z'$ through higher dimension operators - allows much of the usual $Z'$ phenomenology to be realized without spoiling the appealing features of the SM. Such effective $Z'$s can be realized simply through the addition of additional vectorlike matter. Such interactions can address many recent anomalies from the Tevatron, for instance the dimuon asymmetry, the excess of events in the  $Wjj$ search, and the top forward-backward asymmetry (although with tensions in the latter two common to most $Z'$ approaches).

The question remains: how does one {\em search} for an effective $Z'$? While one can search for the dijet or dilepton resonances common to $Z'$, how do we search for a $Z'$ beyond this, or determine that what we have found is coupling to the SM through effective charges?

One possible scenario would be to search for a process by which a $Z'$ was produced through its (dominant) effective couplings, but decays on occasion through an induced kinetic mixing. For instance, by studying the final states, and seeing appropriate ratios for charged and neutral leptons BRs as predicted by (\ref{eq:kinmixingZprime}), but a much stronger production cross section\footnote{or equivalently, a much smaller leptonic BR}, arising from effective quark charges, one could give evidence of this scenario.

\begin{figure}[t] 
   \centering
   \includegraphics[width=0.4\columnwidth]{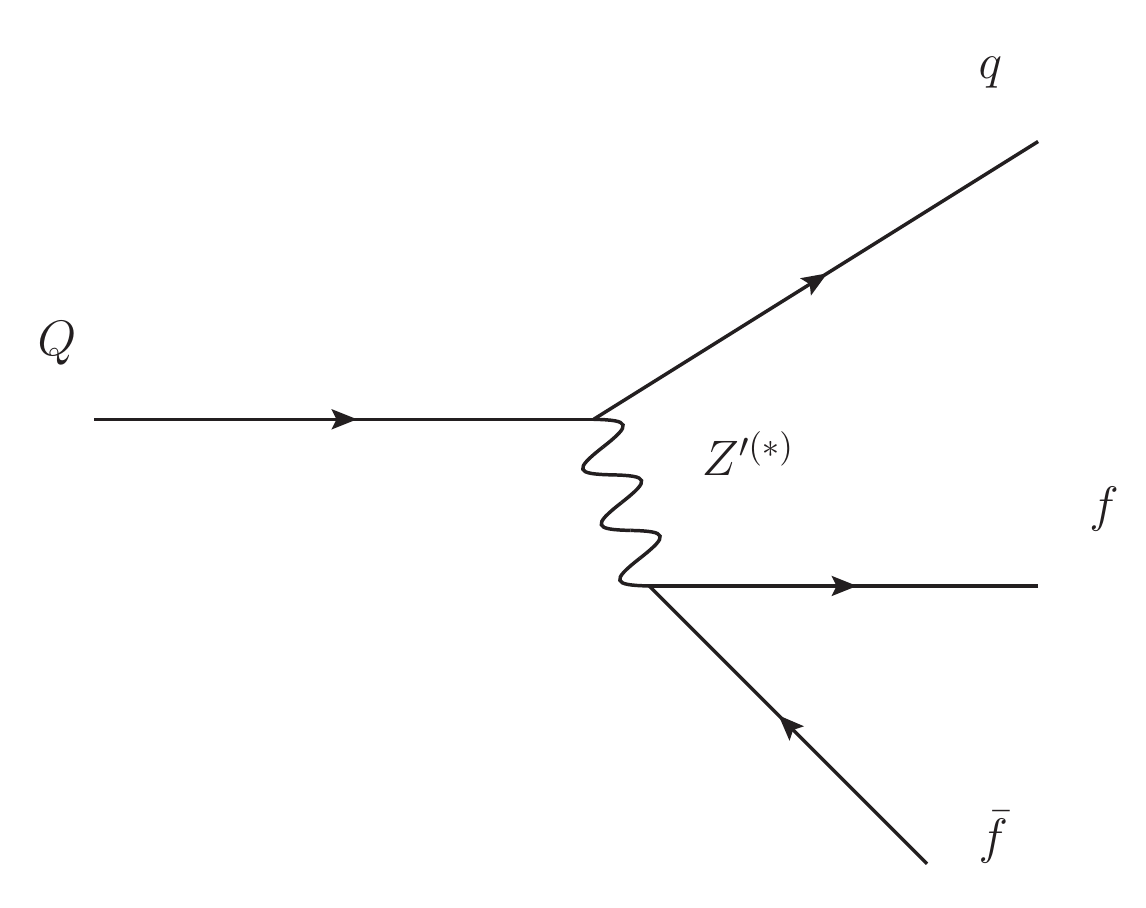} 
   \caption{The decay of a heavy field (shown here as a $Q$) can proceed through off- or on-shell $Z'$ to two body final states.}
   \label{fig:heavyqdecay}
\end{figure}

The most direct way to look test for effective $Z'$s is to look for decays of the states which produce the effective interactions, i.e., the vectorlike fermions, with a fundamental charge under the new group. An example of such a process is shown in Figure~\ref{fig:heavyqdecay}. The decay can proceed through an off- or on-shell $Z'$, but results in interesting three-body final states that reconstruct to the original $Q$ invariant masses. 

%%%%%%%%%
%%%%%%%%%

There is the obvious possibility of three-jet resonances, which if pair-produced through strong interactions will have signatures similar to SUSY with hadronic R-parity violation. Here it is possible for dijets within the trijet resonance to show structure as well. Moreover, unlike SUSY, the heavy $Q$ fields could be singly produced, and may show up in searches for fourth generations, decaying through hadronic channels.

An exciting possibility that arises would be the scenario where the $Z'$ decays leptonically, where final states such as $j l^+l^-$ can reconstruct to invariant masses, even when the dilepton pair does not (i.e., when the $Z'$ is offshell). Thus, to search for effective $Z'$s, one ought to search not only for two body final states (evidence for the $Z'$) but also three body final states (evidence for the physics that produces the effective couplings).

In summary, effective $Z'$ interactions provide a natural way to realize the phenomenology of $Z'$s at the LHC and Tevatron, but without the theoretical problems of conventional $Z'$ models. As the energy and luminosity of the LHC march forward, such scenarios provide an exciting avenue to be pursued.

\section*{Acknowledgements}
The authors thank Wolfgang Altmannshofer, Kyle Cranmer, Bogdan Dobrescu, Roni Harnik, Graham Kribs and Adam Martin for many helpful discussions.   NW is supported by NSF grant PHY-0449818, as well as by the Ambrose Monell Foundation. DTS is is supported by NSF grant PHY-0856522. The authors thank Nima Arkani-Hamed for extensive discussions, and for reminding them that if you haven't understood it in the effective theory, you haven't understood it. The authors thank the IAS for kind hospitality while parts of this work were undertaken.  Fermilab is operated by Fermi Research Alliance, LLC, under Contract DE-AC02-07CH11359 with the United States Department of Energy.

%%%%%%%%%%%%%%%%%%%%
\bibliography{zprimes-allresub}

%Merlin.mbs v4.21 2009-07-09.
\begin{thebibliography}{10}%
\makeatletter
\providecommand \@ifxundefined [1]{%
 \ifx #1\undefined \expandafter \@firstoftwo
 \else \expandafter \@secondoftwo
\fi
}%
\providecommand \@ifnum [1]{%
 \ifnum #1\expandafter \@firstoftwo
 \else \expandafter \@secondoftwo
\fi
}%
\providecommand \enquote [1]{``#1''}%
\providecommand \bibnamefont  [1]{#1}%
\providecommand \bibfnamefont [1]{#1}%
\providecommand \citenamefont [1]{#1}%
\providecommand\href[0]{\@sanitize\@href}%
\providecommand\@href[1]{\endgroup\@@startlink{#1}\endgroup\@@href}%
\providecommand\@@href[1]{#1\@@endlink}%
\providecommand \@sanitize [0]{\begingroup\catcode`\&12\catcode`\#12\relax}%
\@ifxundefined \pdfoutput {\@firstoftwo}{%
 \@ifnum{\z@=\pdfoutput}{\@firstoftwo}{\@secondoftwo}%
}{%
 \providecommand\@@startlink[1]{\leavevmode\special{html:<a href="#1">}}%
 \providecommand\@@endlink[0]{\special{html:</a>}}%
}{%
 \providecommand\@@startlink[1]{%
  \leavevmode
  \pdfstartlink
   attr{/Border[0 0 1 ]/H/I/C[0 1 1]}%
   user{/Subtype/Link/A<</Type/Action/S/URI/URI(#1)>>}%
  \relax
 }%
 \providecommand\@@endlink[0]{\pdfendlink}%
}%
\providecommand \url  [0]{\begingroup\@sanitize \@url }%
\providecommand \@url [1]{\endgroup\@href {#1}{\urlprefix}}%
\providecommand \urlprefix [0]{URL }%
\providecommand \Eprint[0]{\href }%
\@ifxundefined \urlstyle {%
  \providecommand \doi [1]{doi:\discretionary{}{}{}#1}%
}{%
  \providecommand \doi [0]{doi:\discretionary{}{}{}\begingroup
  \urlstyle{rm}\Url }%
}%
\providecommand \doibase [0]{http://dx.doi.org/}%
\providecommand \Doi[1]{\href{\doibase#1}}%
\providecommand \bibAnnote [3]{%
  \BibitemShut{#1}%
  \begin{quotation}\noindent
    \textsc{Key:}\ #2\\\textsc{Annotation:}\ #3%
  \end{quotation}%
}%
\providecommand \bibAnnoteFile [2]{%
  \IfFileExists{#2}{\bibAnnote {#1} {#2} {\input{#2}}}{}%
}%
\providecommand \typeout [0]{\immediate \write \m@ne }%
\providecommand \selectlanguage [0]{\@gobble}%
\providecommand \bibinfo [0]{\@secondoftwo}%
\providecommand \bibfield [0]{\@secondoftwo}%
\providecommand \translation [1]{[#1]}%
\providecommand \BibitemOpen[0]{}%
\providecommand \bibitemStop [0]{}%
\providecommand \bibitemNoStop [0]{.\EOS\space}%
\providecommand \EOS [0]{\spacefactor3000\relax}%
\providecommand \BibitemShut [1]{\csname bibitem#1\endcsname}%
%</preamble>
\bibitem{Kribs:2007nz}%
  \BibitemOpen
  \bibfield{author}{%
  \bibinfo {author} {\bibfnamefont{G.~D.}\ \bibnamefont{Kribs}}, \bibinfo
  {author} {\bibfnamefont{T.}~\bibnamefont{Plehn}}, \bibinfo {author}
  {\bibfnamefont{M.}~\bibnamefont{Spannowsky}},\ and\ \bibinfo {author}
  {\bibfnamefont{T.~M.~P.}\ \bibnamefont{Tait}},\ }%
  \bibfield{journal}{%
  \Doi{10.1103/PhysRevD.76.075016}{\bibinfo {journal} {Phys. Rev.}}\ }%
  \textbf{\bibinfo {volume} {D76}},\ \bibinfo {pages} {075016} (\bibinfo {year}
  {2007}),\ \Eprint{http://arxiv.org/abs/0706.3718}{arXiv:0706.3718 [hep-ph]}%
  \bibAnnoteFile{NoStop}{Kribs:2007nz}%
%%CITATION = 0706.3718;%%
\bibitem{Alcaraz:2006mx}%
  \BibitemOpen
  \bibfield{author}{%
  \bibinfo {author} {\bibfnamefont{J.}~\bibnamefont{Alcaraz}} \emph{et~al.}
  (\bibinfo {collaboration} {ALEPH})}%
   (\bibinfo {year} {2006}),\
  \Eprint{http://arxiv.org/abs/hep-ex/0612034}{arXiv:hep-ex/0612034}%
  \bibAnnoteFile{NoStop}{Alcaraz:2006mx}%
%%CITATION = HEP-EX/0612034;%%
\bibitem{Abazov:2010hv}%
  \BibitemOpen
  \bibfield{author}{%
  \bibinfo {author} {\bibfnamefont{V.~M.}\ \bibnamefont{Abazov}} \emph{et~al.}
  (\bibinfo {collaboration} {D0 Collaboration}),\ }%
  \bibfield{journal}{%
  \Doi{10.1103/PhysRevD.82.032001}{\bibinfo {journal} {Phys.Rev.}}\ }%
  \textbf{\bibinfo {volume} {D82}},\ \bibinfo {pages} {032001} (\bibinfo {year}
  {2010}),\ \Eprint{http://arxiv.org/abs/1005.2757}{arXiv:1005.2757 [hep-ex]}%
  \bibAnnoteFile{NoStop}{Abazov:2010hv}%
\bibitem{Aaltonen:2011mk}%
  \BibitemOpen
  \bibfield{author}{%
  \bibinfo {author} {\bibfnamefont{T.}~\bibnamefont{Aaltonen}} \emph{et~al.}
  (\bibinfo {collaboration} {CDF})}%
   (\bibinfo {year} {2011}),\
  \Eprint{http://arxiv.org/abs/1104.0699}{arXiv:1104.0699 [hep-ex]}%
  \bibAnnoteFile{NoStop}{Aaltonen:2011mk}%
%%CITATION = 1104.0699;%%
\bibitem{Aaltonen:2011kc}%
  \BibitemOpen
  \bibfield{author}{%
  \bibinfo {author} {\bibfnamefont{T.}~\bibnamefont{Aaltonen}} \emph{et~al.}
  (\bibinfo {collaboration} {CDF Collaboration}),\ }%
  \bibfield{journal}{%
  \bibinfo {journal} {Phys.Rev.D}}%
   (\bibinfo {year} {2011}),\
  \Eprint{http://arxiv.org/abs/1101.0034}{arXiv:1101.0034 [hep-ex]}%
  \bibAnnoteFile{NoStop}{Aaltonen:2011kc}%
\bibitem{Langacker:2008yv}%
  \BibitemOpen
  \bibfield{author}{%
  \bibinfo {author} {\bibfnamefont{P.}~\bibnamefont{Langacker}},\ }%
  \bibfield{journal}{%
  \Doi{10.1103/RevModPhys.81.1199}{\bibinfo {journal} {Rev. Mod. Phys.}}\ }%
  \textbf{\bibinfo {volume} {81}},\ \bibinfo {pages} {1199} (\bibinfo {year}
  {2009}),\ \Eprint{http://arxiv.org/abs/0801.1345}{arXiv:0801.1345 [hep-ph]}%
  \bibAnnoteFile{NoStop}{Langacker:2008yv}%
%%CITATION = 0801.1345;%%
\bibitem{Langacker:1988ur}%
  \BibitemOpen
  \bibfield{author}{%
  \bibinfo {author} {\bibfnamefont{P.}~\bibnamefont{Langacker}}\ and\ \bibinfo
  {author} {\bibfnamefont{D.}~\bibnamefont{London}},\ }%
  \bibfield{journal}{%
  \Doi{10.1103/PhysRevD.38.886}{\bibinfo {journal} {Phys.Rev.}}\ }%
  \textbf{\bibinfo {volume} {D38}},\ \bibinfo {pages} {886} (\bibinfo {year}
  {1988})%
  \bibAnnoteFile{NoStop}{Langacker:1988ur}%
\bibitem{Nardi:1992nq}%
  \BibitemOpen
  \bibfield{author}{%
  \bibinfo {author} {\bibfnamefont{E.}~\bibnamefont{Nardi}},\ }%
  \bibfield{journal}{%
  \Doi{10.1103/PhysRevD.48.1240}{\bibinfo {journal} {Phys. Rev.}}\ }%
  \textbf{\bibinfo {volume} {D48}},\ \bibinfo {pages} {1240} (\bibinfo {year}
  {1993}),\ \Eprint{http://arxiv.org/abs/hep-ph/9209223}{arXiv:hep-ph/9209223}%
  \bibAnnoteFile{NoStop}{Nardi:1992nq}%
%%CITATION = HEP-PH/9209223;%%
\bibitem{Nardi:1994iv}%
  \BibitemOpen
  \bibfield{author}{%
  \bibinfo {author} {\bibfnamefont{E.}~\bibnamefont{Nardi}}, \bibinfo {author}
  {\bibfnamefont{E.}~\bibnamefont{Roulet}},\ and\ \bibinfo {author}
  {\bibfnamefont{D.}~\bibnamefont{Tommasini}},\ }%
  \bibfield{journal}{%
  \Doi{10.1016/0370-2693(94)90736-6}{\bibinfo {journal} {Phys.Lett.}}\ }%
  \textbf{\bibinfo {volume} {B327}},\ \bibinfo {pages} {319} (\bibinfo {year}
  {1994}),\ \Eprint{http://arxiv.org/abs/hep-ph/9402224}{arXiv:hep-ph/9402224
  [hep-ph]}%
  \bibAnnoteFile{NoStop}{Nardi:1994iv}%
\bibitem{ArkaniHamed:2001nc}%
  \BibitemOpen
  \bibfield{author}{%
  \bibinfo {author} {\bibfnamefont{N.}~\bibnamefont{Arkani-Hamed}}, \bibinfo
  {author} {\bibfnamefont{A.~G.}\ \bibnamefont{Cohen}},\ and\ \bibinfo {author}
  {\bibfnamefont{H.}~\bibnamefont{Georgi}},\ }%
  \bibfield{journal}{%
  \Doi{10.1016/S0370-2693(01)00741-9}{\bibinfo {journal} {Phys. Lett.}}\ }%
  \textbf{\bibinfo {volume} {B513}},\ \bibinfo {pages} {232} (\bibinfo {year}
  {2001}),\ \Eprint{http://arxiv.org/abs/hep-ph/0105239}{arXiv:hep-ph/0105239}%
  \bibAnnoteFile{NoStop}{ArkaniHamed:2001nc}%
%%CITATION = HEP-PH/0105239;%%
\bibitem{Buckley:2011vc}%
  \BibitemOpen
  \bibfield{author}{%
  \bibinfo {author} {\bibfnamefont{M.~R.}\ \bibnamefont{Buckley}}, \bibinfo
  {author} {\bibfnamefont{D.}~\bibnamefont{Hooper}}, \bibinfo {author}
  {\bibfnamefont{J.}~\bibnamefont{Kopp}},\ and\ \bibinfo {author}
  {\bibfnamefont{E.}~\bibnamefont{Neil}}}%
   (\bibinfo {year} {2011}),\
  \Eprint{http://arxiv.org/abs/1103.6035}{arXiv:1103.6035 [hep-ph]}%
  \bibAnnoteFile{NoStop}{Buckley:2011vc}%
%%CITATION = 1103.6035;%%
\bibitem{Nierste:2011ti}%
  \BibitemOpen
  \bibfield{author}{%
  \bibinfo {author} {\bibfnamefont{A.~L.}\ \bibnamefont{Nierste}}\ and\
  \bibinfo {author} {\bibnamefont{Ulrich}}}%
   (\bibinfo {year} {2011}),\
  \Eprint{http://arxiv.org/abs/1102.4274}{arXiv:1102.4274 [hep-ph]}%
  \bibAnnoteFile{NoStop}{Nierste:2011ti}%
\bibitem{Lenz:2006hd}%
  \BibitemOpen
  \bibfield{author}{%
  \bibinfo {author} {\bibfnamefont{A.}~\bibnamefont{Lenz}}\ and\ \bibinfo
  {author} {\bibfnamefont{U.}~\bibnamefont{Nierste}},\ }%
  \bibfield{journal}{%
  \Doi{10.1088/1126-6708/2007/06/072}{\bibinfo {journal} {JHEP}}\ }%
  \textbf{\bibinfo {volume} {06}},\ \bibinfo {pages} {072} (\bibinfo {year}
  {2007}),\ \Eprint{http://arxiv.org/abs/hep-ph/0612167}{arXiv:hep-ph/0612167}%
  \bibAnnoteFile{NoStop}{Lenz:2006hd}%
%%CITATION = HEP-PH/0612167;%%
\bibitem{cdfAslnote}%
  \BibitemOpen
  \bibfield{author}{%
  \bibinfo {author} {\bibnamefont{{The CDF Collaboration}}},\ }%
  \bibinfo {howpublished} {CDF Note 9015} (\bibinfo {year} {2007})%
  \bibAnnoteFile{NoStop}{cdfAslnote}%
\bibitem{cdfd0combinedBs}%
  \BibitemOpen
  \bibfield{author}{%
  \bibinfo {author} {\bibnamefont{{The CDF and D0 Collaborations}}},\ }%
  \bibinfo {howpublished} {CDF Note 9787/D0 5928-CONF} (\bibinfo {year}
  {2009})%
  \bibAnnoteFile{NoStop}{cdfd0combinedBs}%
\bibitem{cdfBs2010}%
  \BibitemOpen
  \bibfield{author}{%
  \bibinfo {author} {\bibnamefont{{The CDF Collaboration}}},\ }%
  \bibinfo {howpublished} {CDF Note 10206} (\bibinfo {year} {2010})%
  \bibAnnoteFile{NoStop}{cdfBs2010}%
\bibitem{d0Bs2010}%
  \BibitemOpen
  \bibfield{author}{%
  \bibinfo {author} {\bibnamefont{{The CDF Collaboration}}},\ }%
  \bibinfo {howpublished} {D0 Note 6098-CONF} (\bibinfo {year} {2010})%
  \bibAnnoteFile{NoStop}{d0Bs2010}%
\bibitem{Lenz:2010gu}%
  \BibitemOpen
  \bibfield{author}{%
  \bibinfo {author} {\bibfnamefont{A.}~\bibnamefont{Lenz}} \emph{et~al.},\ }%
  \bibfield{journal}{%
  \Doi{10.1103/PhysRevD.83.036004}{\bibinfo {journal} {Phys. Rev.}}\ }%
  \textbf{\bibinfo {volume} {D83}},\ \bibinfo {pages} {036004} (\bibinfo {year}
  {2011}),\ \Eprint{http://arxiv.org/abs/1008.1593}{arXiv:1008.1593 [hep-ph]}%
  \bibAnnoteFile{NoStop}{Lenz:2010gu}%
%%CITATION = 1008.1593;%%
\bibitem{Cheung:2006tm}%
  \BibitemOpen
  \bibfield{author}{%
  \bibinfo {author} {\bibfnamefont{K.}~\bibnamefont{Cheung}}, \bibinfo {author}
  {\bibfnamefont{C.-W.}\ \bibnamefont{Chiang}}, \bibinfo {author}
  {\bibfnamefont{N.~G.}\ \bibnamefont{Deshpande}},\ and\ \bibinfo {author}
  {\bibfnamefont{J.}~\bibnamefont{Jiang}},\ }%
  \bibfield{journal}{%
  \Doi{10.1016/j.physletb.2007.07.032}{\bibinfo {journal} {Phys. Lett.}}\ }%
  \textbf{\bibinfo {volume} {B652}},\ \bibinfo {pages} {285} (\bibinfo {year}
  {2007}),\ \Eprint{http://arxiv.org/abs/hep-ph/0604223}{arXiv:hep-ph/0604223}%
  \bibAnnoteFile{NoStop}{Cheung:2006tm}%
%%CITATION = HEP-PH/0604223;%%
\bibitem{Chiang:2006we}%
  \BibitemOpen
  \bibfield{author}{%
  \bibinfo {author} {\bibfnamefont{C.-W.}\ \bibnamefont{Chiang}}, \bibinfo
  {author} {\bibfnamefont{N.~G.}\ \bibnamefont{Deshpande}},\ and\ \bibinfo
  {author} {\bibfnamefont{J.}~\bibnamefont{Jiang}},\ }%
  \bibfield{journal}{%
  \Doi{10.1088/1126-6708/2006/08/075}{\bibinfo {journal} {JHEP}}\ }%
  \textbf{\bibinfo {volume} {08}},\ \bibinfo {pages} {075} (\bibinfo {year}
  {2006}),\ \Eprint{http://arxiv.org/abs/hep-ph/0606122}{arXiv:hep-ph/0606122}%
  \bibAnnoteFile{NoStop}{Chiang:2006we}%
%%CITATION = HEP-PH/0606122;%%
\bibitem{He:2009ie}%
  \BibitemOpen
  \bibfield{author}{%
  \bibinfo {author} {\bibfnamefont{X.-G.}\ \bibnamefont{He}}\ and\ \bibinfo
  {author} {\bibfnamefont{G.}~\bibnamefont{Valencia}},\ }%
  \bibfield{journal}{%
  \Doi{10.1016/j.physletb.2009.08.033}{\bibinfo {journal} {Phys. Lett.}}\ }%
  \textbf{\bibinfo {volume} {B680}},\ \bibinfo {pages} {72} (\bibinfo {year}
  {2009}),\ \Eprint{http://arxiv.org/abs/0907.4034}{arXiv:0907.4034 [hep-ph]}%
  \bibAnnoteFile{NoStop}{He:2009ie}%
%%CITATION = 0907.4034;%%
\bibitem{Barger:2009eq}%
  \BibitemOpen
  \bibfield{author}{%
  \bibinfo {author} {\bibfnamefont{V.}~\bibnamefont{Barger}} \emph{et~al.},\ }%
  \bibfield{journal}{%
  \Doi{10.1103/PhysRevD.80.055008}{\bibinfo {journal} {Phys. Rev.}}\ }%
  \textbf{\bibinfo {volume} {D80}},\ \bibinfo {pages} {055008} (\bibinfo {year}
  {2009}),\ \Eprint{http://arxiv.org/abs/0902.4507}{arXiv:0902.4507 [hep-ph]}%
  \bibAnnoteFile{NoStop}{Barger:2009eq}%
%%CITATION = 0902.4507;%%
\bibitem{Barger:2009qs}%
  \BibitemOpen
  \bibfield{author}{%
  \bibinfo {author} {\bibfnamefont{V.}~\bibnamefont{Barger}} \emph{et~al.},\ }%
  \bibfield{journal}{%
  \Doi{10.1088/1126-6708/2009/12/048}{\bibinfo {journal} {JHEP}}\ }%
  \textbf{\bibinfo {volume} {12}},\ \bibinfo {pages} {048} (\bibinfo {year}
  {2009}),\ \Eprint{http://arxiv.org/abs/0906.3745}{arXiv:0906.3745 [hep-ph]}%
  \bibAnnoteFile{NoStop}{Barger:2009qs}%
%%CITATION = 0906.3745;%%
\bibitem{Deshpande:2010hy}%
  \BibitemOpen
  \bibfield{author}{%
  \bibinfo {author} {\bibfnamefont{N.~G.}\ \bibnamefont{Deshpande}}, \bibinfo
  {author} {\bibfnamefont{X.-G.}\ \bibnamefont{He}},\ and\ \bibinfo {author}
  {\bibfnamefont{G.}~\bibnamefont{Valencia}},\ }%
  \bibfield{journal}{%
  \Doi{10.1103/PhysRevD.82.056013}{\bibinfo {journal} {Phys. Rev.}}\ }%
  \textbf{\bibinfo {volume} {D82}},\ \bibinfo {pages} {056013} (\bibinfo {year}
  {2010}),\ \Eprint{http://arxiv.org/abs/1006.1682}{arXiv:1006.1682 [hep-ph]}%
  \bibAnnoteFile{NoStop}{Deshpande:2010hy}%
%%CITATION = 1006.1682;%%
\bibitem{Alok:2010ij}%
  \BibitemOpen
  \bibfield{author}{%
  \bibinfo {author} {\bibfnamefont{A.~K.}\ \bibnamefont{Alok}}, \bibinfo
  {author} {\bibfnamefont{S.}~\bibnamefont{Baek}},\ and\ \bibinfo {author}
  {\bibfnamefont{D.}~\bibnamefont{London}}}%
   (\bibinfo {year} {2010}),\
  \Eprint{http://arxiv.org/abs/1010.1333}{arXiv:1010.1333 [hep-ph]}%
  \bibAnnoteFile{NoStop}{Alok:2010ij}%
%%CITATION = 1010.1333;%%
\bibitem{Kim:2010gx}%
  \BibitemOpen
  \bibfield{author}{%
  \bibinfo {author} {\bibfnamefont{J.~E.}\ \bibnamefont{Kim}}, \bibinfo
  {author} {\bibfnamefont{M.-S.}\ \bibnamefont{Seo}},\ and\ \bibinfo {author}
  {\bibfnamefont{S.}~\bibnamefont{Shin}},\ }%
  \bibfield{journal}{%
  \Doi{10.1103/PhysRevD.83.036003}{\bibinfo {journal} {Phys. Rev.}}\ }%
  \textbf{\bibinfo {volume} {D83}},\ \bibinfo {pages} {036003} (\bibinfo {year}
  {2011}),\ \Eprint{http://arxiv.org/abs/1010.5123}{arXiv:1010.5123 [hep-ph]}%
  \bibAnnoteFile{NoStop}{Kim:2010gx}%
%%CITATION = 1010.5123;%%
\bibitem{Buras:1990fn}%
  \BibitemOpen
  \bibfield{author}{%
  \bibinfo {author} {\bibfnamefont{A.~J.}\ \bibnamefont{Buras}}, \bibinfo
  {author} {\bibfnamefont{M.}~\bibnamefont{Jamin}},\ and\ \bibinfo {author}
  {\bibfnamefont{P.~H.}\ \bibnamefont{Weisz}},\ }%
  \bibfield{journal}{%
  \Doi{10.1016/0550-3213(90)90373-L}{\bibinfo {journal} {Nucl.Phys.}}\ }%
  \textbf{\bibinfo {volume} {B347}},\ \bibinfo {pages} {491} (\bibinfo {year}
  {1990})%
  \bibAnnoteFile{NoStop}{Buras:1990fn}%
\bibitem{Abulencia:2006ze}%
  \BibitemOpen
  \bibfield{author}{%
  \bibinfo {author} {\bibfnamefont{A.}~\bibnamefont{Abulencia}} \emph{et~al.}
  (\bibinfo {collaboration} {CDF Collaboration}),\ }%
  \bibfield{journal}{%
  \Doi{10.1103/PhysRevLett.97.242003}{\bibinfo {journal} {Phys.Rev.Lett.}}\ }%
  \textbf{\bibinfo {volume} {97}},\ \bibinfo {pages} {242003} (\bibinfo {year}
  {2006}),\ \Eprint{http://arxiv.org/abs/hep-ex/0609040}{arXiv:hep-ex/0609040
  [hep-ex]}%
  \bibAnnoteFile{NoStop}{Abulencia:2006ze}%
\bibitem{Nakamura:2010zz}%
  \BibitemOpen
  \bibfield{author}{%
  \bibinfo {author} {\bibfnamefont{K.}~\bibnamefont{Nakamura}} \emph{et~al.}
  (\bibinfo {collaboration} {Particle Data Group}),\ }%
  \bibfield{journal}{%
  \Doi{10.1088/0954-3899/37/7A/075021}{\bibinfo {journal} {J.Phys.G}}\ }%
  \textbf{\bibinfo {volume} {G37}},\ \bibinfo {pages} {075021} (\bibinfo {year}
  {2010})%
  \bibAnnoteFile{NoStop}{Nakamura:2010zz}%
\bibitem{Isidori:2010kg}%
  \BibitemOpen
  \bibfield{author}{%
  \bibinfo {author} {\bibfnamefont{G.}~\bibnamefont{Isidori}}, \bibinfo
  {author} {\bibfnamefont{Y.}~\bibnamefont{Nir}},\ and\ \bibinfo {author}
  {\bibfnamefont{G.}~\bibnamefont{Perez}}}%
   (\bibinfo {year} {2010}),\
  \Eprint{http://arxiv.org/abs/1002.0900}{arXiv:1002.0900 [hep-ph]}%
  \bibAnnoteFile{NoStop}{Isidori:2010kg}%
\bibitem{Aubert:2008rr}%
  \BibitemOpen
  \bibfield{author}{%
  \bibinfo {author} {\bibfnamefont{B.}~\bibnamefont{Aubert}} \emph{et~al.}
  (\bibinfo {collaboration} {BABAR Collaboration}),\ }%
  \bibfield{journal}{%
  \Doi{10.1103/PhysRevD.78.091102}{\bibinfo {journal} {Phys.Rev.}}\ }%
  \textbf{\bibinfo {volume} {D78}},\ \bibinfo {pages} {091102} (\bibinfo {year}
  {2008}),\ \Eprint{http://arxiv.org/abs/0808.0900}{arXiv:0808.0900 [hep-ex]}%
  \bibAnnoteFile{NoStop}{Aubert:2008rr}%
\bibitem{Fajfer:2001te}%
  \BibitemOpen
  \bibfield{author}{%
  \bibinfo {author} {\bibfnamefont{S.}~\bibnamefont{Fajfer}}\ and\ \bibinfo
  {author} {\bibfnamefont{P.}~\bibnamefont{Singer}},\ }%
  \bibfield{journal}{%
  \Doi{10.1016/S0920-5632(00)01070-7}{\bibinfo {journal} {Nucl. Phys. Proc.
  Suppl.}}\ }%
  \textbf{\bibinfo {volume} {93}},\ \bibinfo {pages} {103} (\bibinfo {year}
  {2001}),\ \Eprint{http://arxiv.org/abs/hep-ph/0102135}{arXiv:hep-ph/0102135}%
  \bibAnnoteFile{NoStop}{Fajfer:2001te}%
%%CITATION = HEP-PH/0102135;%%
\bibitem{Eichten:2011sh}%
  \BibitemOpen
  \bibfield{author}{%
  \bibinfo {author} {\bibfnamefont{E.~J.}\ \bibnamefont{Eichten}}, \bibinfo
  {author} {\bibfnamefont{K.}~\bibnamefont{Lane}},\ and\ \bibinfo {author}
  {\bibfnamefont{A.}~\bibnamefont{Martin}}}%
   (\bibinfo {year} {2011}),\
  \Eprint{http://arxiv.org/abs/1104.0976}{arXiv:1104.0976 [hep-ph]}%
  \bibAnnoteFile{NoStop}{Eichten:2011sh}%
%%CITATION = 1104.0976;%%
\bibitem{Yu:2011cw}%
  \BibitemOpen
  \bibfield{author}{%
  \bibinfo {author} {\bibfnamefont{F.}~\bibnamefont{Yu}}}%
   (\bibinfo {year} {2011}),\
  \Eprint{http://arxiv.org/abs/1104.0243}{arXiv:1104.0243 [hep-ph]}%
  \bibAnnoteFile{NoStop}{Yu:2011cw}%
%%CITATION = 1104.0243;%%
\bibitem{Kilic:2011sr}%
  \BibitemOpen
  \bibfield{author}{%
  \bibinfo {author} {\bibfnamefont{C.}~\bibnamefont{Kilic}}\ and\ \bibinfo
  {author} {\bibfnamefont{S.}~\bibnamefont{Thomas}}}%
   (\bibinfo {year} {2011}),\
  \Eprint{http://arxiv.org/abs/1104.1002}{arXiv:1104.1002 [hep-ph]}%
  \bibAnnoteFile{NoStop}{Kilic:2011sr}%
\bibitem{AguilarSaavedra:2011zy}%
  \BibitemOpen
  \bibfield{author}{%
  \bibinfo {author} {\bibfnamefont{J.}~\bibnamefont{Aguilar-Saavedra}}\ and\
  \bibinfo {author} {\bibfnamefont{M.}~\bibnamefont{Perez-Victoria}}}%
   (\bibinfo {year} {2011}),\
  \Eprint{http://arxiv.org/abs/1104.1385}{arXiv:1104.1385 [hep-ph]}%
  \bibAnnoteFile{NoStop}{AguilarSaavedra:2011zy}%
\bibitem{Cheung:2011zt}%
  \BibitemOpen
  \bibfield{author}{%
  \bibinfo {author} {\bibfnamefont{K.}~\bibnamefont{Cheung}}\ and\ \bibinfo
  {author} {\bibfnamefont{J.}~\bibnamefont{Song}}}%
   (\bibinfo {year} {2011}),\
  \Eprint{http://arxiv.org/abs/1104.1375}{arXiv:1104.1375 [hep-ph]}%
  \bibAnnoteFile{NoStop}{Cheung:2011zt}%
\bibitem{Nelson:2011us}%
  \BibitemOpen
  \bibfield{author}{%
  \bibinfo {author} {\bibfnamefont{A.~E.}\ \bibnamefont{Nelson}}, \bibinfo
  {author} {\bibfnamefont{T.}~\bibnamefont{Okui}},\ and\ \bibinfo {author}
  {\bibfnamefont{T.~S.}\ \bibnamefont{Roy}}}%
   (\bibinfo {year} {2011}),\
  \Eprint{http://arxiv.org/abs/1104.2030}{arXiv:1104.2030 [hep-ph]}%
  \bibAnnoteFile{NoStop}{Nelson:2011us}%
\bibitem{Sato:2011ui}%
  \BibitemOpen
  \bibfield{author}{%
  \bibinfo {author} {\bibfnamefont{R.}~\bibnamefont{Sato}}, \bibinfo {author}
  {\bibfnamefont{S.}~\bibnamefont{Shirai}},\ and\ \bibinfo {author}
  {\bibfnamefont{K.}~\bibnamefont{Yonekura}}}%
   (\bibinfo {year} {2011}),\
  \Eprint{http://arxiv.org/abs/1104.2014}{arXiv:1104.2014 [hep-ph]}%
  \bibAnnoteFile{NoStop}{Sato:2011ui}%
\bibitem{Wang:2011ta}%
  \BibitemOpen
  \bibfield{author}{%
  \bibinfo {author} {\bibfnamefont{X.-P.}\ \bibnamefont{Wang}}, \bibinfo
  {author} {\bibfnamefont{Y.-K.}\ \bibnamefont{Wang}}, \bibinfo {author}
  {\bibfnamefont{B.}~\bibnamefont{Xiao}}, \bibinfo {author}
  {\bibfnamefont{J.}~\bibnamefont{Xu}},\ and\ \bibinfo {author}
  {\bibfnamefont{S.-h.}\ \bibnamefont{Zhu}}}%
   (\bibinfo {year} {2011}),\
  \Eprint{http://arxiv.org/abs/1104.1917}{arXiv:1104.1917 [hep-ph]}%
  \bibAnnoteFile{NoStop}{Wang:2011ta}%
\bibitem{He:2011ss}%
  \BibitemOpen
  \bibfield{author}{%
  \bibinfo {author} {\bibfnamefont{X.-G.}\ \bibnamefont{He}}\ and\ \bibinfo
  {author} {\bibfnamefont{B.-Q.}\ \bibnamefont{Ma}}}%
   (\bibinfo {year} {2011}),\
  \Eprint{http://arxiv.org/abs/1104.1894}{arXiv:1104.1894 [hep-ph]}%
  \bibAnnoteFile{NoStop}{He:2011ss}%
\bibitem{Anchordoqui:2011ag}%
  \BibitemOpen
  \bibfield{author}{%
  \bibinfo {author} {\bibfnamefont{L.~A.}\ \bibnamefont{Anchordoqui}}, \bibinfo
  {author} {\bibfnamefont{H.}~\bibnamefont{Goldberg}}, \bibinfo {author}
  {\bibfnamefont{X.}~\bibnamefont{Huang}}, \bibinfo {author}
  {\bibfnamefont{D.}~\bibnamefont{Lust}},\ and\ \bibinfo {author}
  {\bibfnamefont{T.~R.}\ \bibnamefont{Taylor}}}%
   (\bibinfo {year} {2011}),\
  \Eprint{http://arxiv.org/abs/1104.2302}{arXiv:1104.2302 [hep-ph]}%
  \bibAnnoteFile{NoStop}{Anchordoqui:2011ag}%
\bibitem{Dobrescu:2011px}%
  \BibitemOpen
  \bibfield{author}{%
  \bibinfo {author} {\bibfnamefont{B.~A.}\ \bibnamefont{Dobrescu}}\ and\
  \bibinfo {author} {\bibfnamefont{G.~Z.}\ \bibnamefont{Krnjaic}}}%
   (\bibinfo {year} {2011}),\
  \Eprint{http://arxiv.org/abs/1104.2893}{arXiv:1104.2893 [hep-ph]}%
  \bibAnnoteFile{NoStop}{Dobrescu:2011px}%
\bibitem{Fodor:2011tu}%
  \BibitemOpen
  \bibfield{author}{%
  \bibinfo {author} {\bibfnamefont{Z.}~\bibnamefont{Fodor}}, \bibinfo {author}
  {\bibfnamefont{K.}~\bibnamefont{Holland}}, \bibinfo {author}
  {\bibfnamefont{J.}~\bibnamefont{Kuti}}, \bibinfo {author}
  {\bibfnamefont{D.}~\bibnamefont{Nogradi}},\ and\ \bibinfo {author}
  {\bibfnamefont{C.}~\bibnamefont{Schroeder}}}%
   (\bibinfo {year} {2011}),\
  \Eprint{http://arxiv.org/abs/1104.3124}{arXiv:1104.3124 [hep-lat]}%
  \bibAnnoteFile{NoStop}{Fodor:2011tu}%
\bibitem{Buckley:2011vs}%
  \BibitemOpen
  \bibfield{author}{%
  \bibinfo {author} {\bibfnamefont{M.}~\bibnamefont{Buckley}}, \bibinfo
  {author} {\bibfnamefont{P.}~\bibnamefont{Fileviez~Perez}}, \bibinfo {author}
  {\bibfnamefont{D.}~\bibnamefont{Hooper}},\ and\ \bibinfo {author}
  {\bibfnamefont{E.}~\bibnamefont{Neil}}}%
   (\bibinfo {year} {2011}),\
  \Eprint{http://arxiv.org/abs/1104.3145}{arXiv:1104.3145 [hep-ph]}%
  \bibAnnoteFile{NoStop}{Buckley:2011vs}%
%%CITATION = 1104.3145;%%
\bibitem{Alwall:2007st}%
  \BibitemOpen
  \bibfield{author}{%
  \bibinfo {author} {\bibfnamefont{J.}~\bibnamefont{Alwall}}, \bibinfo {author}
  {\bibfnamefont{P.}~\bibnamefont{Demin}}, \bibinfo {author}
  {\bibfnamefont{S.}~\bibnamefont{de~Visscher}}, \bibinfo {author}
  {\bibfnamefont{R.}~\bibnamefont{Frederix}}, \bibinfo {author}
  {\bibfnamefont{M.}~\bibnamefont{Herquet}}, \emph{et~al.},\ }%
  \bibfield{journal}{%
  \Doi{10.1088/1126-6708/2007/09/028}{\bibinfo {journal} {JHEP}}\ }%
  \textbf{\bibinfo {volume} {0709}},\ \bibinfo {pages} {028} (\bibinfo {year}
  {2007}),\ \Eprint{http://arxiv.org/abs/0706.2334}{arXiv:0706.2334 [hep-ph]}%
  \bibAnnoteFile{NoStop}{Alwall:2007st}%
\bibitem{Pumplin:2002vw}%
  \BibitemOpen
  \bibfield{author}{%
  \bibinfo {author} {\bibfnamefont{J.}~\bibnamefont{Pumplin}} \emph{et~al.},\
  }%
  \bibfield{journal}{%
  \bibinfo {journal} {JHEP}\ }%
  \textbf{\bibinfo {volume} {07}},\ \bibinfo {pages} {012} (\bibinfo {year}
  {2002}),\ \Eprint{http://arxiv.org/abs/hep-ph/0201195}{arXiv:hep-ph/0201195}%
  \bibAnnoteFile{NoStop}{Pumplin:2002vw}%
%%CITATION = HEP-PH/0201195;%%
\bibitem{Alitti:1993pn}%
  \BibitemOpen
  \bibfield{author}{%
  \bibinfo {author} {\bibfnamefont{J.}~\bibnamefont{Alitti}} \emph{et~al.}
  (\bibinfo {collaboration} {UA2 Collaboration}),\ }%
  \bibfield{journal}{%
  \Doi{10.1016/0550-3213(93)90395-6}{\bibinfo {journal} {Nucl.Phys.}}\ }%
  \textbf{\bibinfo {volume} {B400}},\ \bibinfo {pages} {3} (\bibinfo {year}
  {1993})%
  \bibAnnoteFile{NoStop}{Alitti:1993pn}%
\bibitem{Aaltonen:2008dn}%
  \BibitemOpen
  \bibfield{author}{%
  \bibinfo {author} {\bibfnamefont{T.}~\bibnamefont{Aaltonen}} \emph{et~al.}
  (\bibinfo {collaboration} {CDF}),\ }%
  \bibfield{journal}{%
  \Doi{10.1103/PhysRevD.79.112002}{\bibinfo {journal} {Phys. Rev.}}\ }%
  \textbf{\bibinfo {volume} {D79}},\ \bibinfo {pages} {112002} (\bibinfo {year}
  {2009}),\ \Eprint{http://arxiv.org/abs/0812.4036}{arXiv:0812.4036 [hep-ex]}%
  \bibAnnoteFile{NoStop}{Aaltonen:2008dn}%
%%CITATION = 0812.4036;%%
\bibitem{Aaltonen:2009fd}%
  \BibitemOpen
  \bibfield{author}{%
  \bibinfo {author} {\bibfnamefont{T.}~\bibnamefont{Aaltonen}} \emph{et~al.}
  (\bibinfo {collaboration} {CDF}),\ }%
  \bibfield{journal}{%
  \Doi{10.1103/PhysRevLett.103.091803}{\bibinfo {journal} {Phys.Rev.Lett.}}\ }%
  \textbf{\bibinfo {volume} {103}},\ \bibinfo {pages} {091803} (\bibinfo {year}
  {2009}),\ \Eprint{http://arxiv.org/abs/0905.4714}{arXiv:0905.4714 [hep-ex]}%
  \bibAnnoteFile{NoStop}{Aaltonen:2009fd}%
\bibitem{cdfphotonjets}%
  \BibitemOpen
  \bibfield{author}{%
  \bibinfo {author} {\bibnamefont{{The CDF Collaboration}}},\ }%
  \bibinfo {howpublished}
  {\url{http://www-cdf.fnal.gov/physics/exotic/r2a/20110203.gammajets/cdf10355%
_AnomalousPhoJetsMet.pdf}} (\bibinfo {year} {2010})%
  \bibAnnoteFile{NoStop}{cdfphotonjets}%
\bibitem{cdfthesis}%
  \BibitemOpen
  \bibfield{author}{%
  \bibinfo {author} {\bibfnamefont{V.}~\bibnamefont{Cavaliere}},\ }%
  \bibinfo {howpublished} {Ph.D. Thesis. University of Siena,
  FERMILAB-THESIS-2010-51} (\bibinfo {year} {2010})%
  \bibAnnoteFile{NoStop}{cdfthesis}%
\bibitem{cdfwebmet}%
  \BibitemOpen
  \bibfield{author}{%
  \bibinfo {author} {\bibnamefont{{The CDF Collaboration}}},\ }%
  \bibinfo {howpublished}
  {\url{http://www-cdf.fnal.gov/physics/new/hdg/results/VVmetjj_090528/Diboson%
s_METJJ.html}} (\bibinfo {year} {2009})%
  \bibAnnoteFile{NoStop}{cdfwebmet}%
\bibitem{Note1}%
  \BibitemOpen
  \bibinfo {note} {We thank Matt Reece for bringing this to our attention.}%
  \bibAnnoteFile{Stop}{Note1}%
\bibitem{gammamet}%
  \BibitemOpen
  \bibfield{author}{%
  \bibinfo {author} {\bibnamefont{{The CDF Collaboration}}},\ }%
  \bibinfo {howpublished}
  {\url{http://www-cdf.fnal.gov/physics/exotic/r2a/20071213.gammamet/LonelyPho%
tons/photonmet.html}} (\bibinfo {year} {2007})%
  \bibAnnoteFile{NoStop}{gammamet}%
\bibitem{privatecommunication}%
  \BibitemOpen
  \bibinfo {howpublished} {{Samantha Hewamanage, private communication.}}%
  \bibAnnoteFile{Stop}{privatecommunication}%
\bibitem{Jung:2011ua}%
  \BibitemOpen
  \bibfield{author}{%
  \bibinfo {author} {\bibfnamefont{S.}~\bibnamefont{Jung}}, \bibinfo {author}
  {\bibfnamefont{A.}~\bibnamefont{Pierce}},\ and\ \bibinfo {author}
  {\bibfnamefont{J.~D.}\ \bibnamefont{Wells}}}%
   (\bibinfo {year} {2011}),\
  \Eprint{http://arxiv.org/abs/1104.3139}{arXiv:1104.3139 [hep-ph]}%
  \bibAnnoteFile{NoStop}{Jung:2011ua}%
\bibitem{Abazov:2011mi}%
  \BibitemOpen
  \bibfield{author}{%
  \bibinfo {author} {\bibfnamefont{V.~M.}\ \bibnamefont{Abazov}} \emph{et~al.}
  (\bibinfo {collaboration} {D0 Collaboration})}%
   (\bibinfo {year} {2011}),\
  \Eprint{http://arxiv.org/abs/1101.0124}{arXiv:1101.0124 [hep-ex]}%
  \bibAnnoteFile{NoStop}{Abazov:2011mi}%
\bibitem{cdfttxsec}%
  \BibitemOpen
  \bibfield{author}{%
  \bibinfo {author} {\bibnamefont{{The CDF Collaboration}}},\ }%
  \bibinfo {howpublished}
  {\url{http://www-cdf.fnal.gov/physics/new/top/confNotes/cdf9913_ttbarxs4invf%
b.ps}} (\bibinfo {year} {2009})%
  \bibAnnoteFile{NoStop}{cdfttxsec}%
\bibitem{Langenfeld:2009wd}%
  \BibitemOpen
  \bibfield{author}{%
  \bibinfo {author} {\bibfnamefont{U.}~\bibnamefont{Langenfeld}}, \bibinfo
  {author} {\bibfnamefont{S.}~\bibnamefont{Moch}},\ and\ \bibinfo {author}
  {\bibfnamefont{P.}~\bibnamefont{Uwer}},\ }%
  \bibfield{journal}{%
  \Doi{10.1103/PhysRevD.80.054009}{\bibinfo {journal} {Phys.Rev.}}\ }%
  \textbf{\bibinfo {volume} {D80}},\ \bibinfo {pages} {054009} (\bibinfo {year}
  {2009}),\ \Eprint{http://arxiv.org/abs/0906.5273}{arXiv:0906.5273 [hep-ph]}%
  \bibAnnoteFile{NoStop}{Langenfeld:2009wd}%
\bibitem{cmssingletop}%
  \BibitemOpen
  \bibfield{author}{%
  \bibinfo {author} {\bibnamefont{{The CMS Collaboration}}},\ }%
  \bibinfo {howpublished} {CMS Physics Analysis Summary TOP 10-008} (\bibinfo
  {year} {2011})%
  \bibAnnoteFile{NoStop}{cmssingletop}%
\bibitem{Group:2009qk}%
  \BibitemOpen
  \bibfield{author}{%
  \bibinfo {author} {\bibnamefont{{Tevatron Electroweak Working Group}}}
  (\bibinfo {collaboration} {CDF and D0})}%
   (\bibinfo {year} {2009}),\
  \Eprint{http://arxiv.org/abs/0908.2171}{arXiv:0908.2171 [hep-ex]}%
  \bibAnnoteFile{NoStop}{Group:2009qk}%
%%CITATION = 0908.2171;%%
\bibitem{Aaltonen:2008hx}%
  \BibitemOpen
  \bibfield{author}{%
  \bibinfo {author} {\bibfnamefont{T.}~\bibnamefont{Aaltonen}} \emph{et~al.}
  (\bibinfo {collaboration} {CDF}),\ }%
  \bibfield{journal}{%
  \Doi{10.1103/PhysRevLett.102.041801}{\bibinfo {journal} {Phys. Rev. Lett.}}\
  }%
  \textbf{\bibinfo {volume} {102}},\ \bibinfo {pages} {041801} (\bibinfo {year}
  {2009}),\ \Eprint{http://arxiv.org/abs/0809.4903}{arXiv:0809.4903 [hep-ex]}%
  \bibAnnoteFile{NoStop}{Aaltonen:2008hx}%
%%CITATION = 0809.4903;%%
\bibitem{Aaltonen:2009nr}%
  \BibitemOpen
  \bibfield{author}{%
  \bibinfo {author} {\bibfnamefont{T.}~\bibnamefont{Aaltonen}} \emph{et~al.}
  (\bibinfo {collaboration} {CDF Collaboration}),\ }%
  \bibfield{journal}{%
  \Doi{10.1103/PhysRevLett.104.091801}{\bibinfo {journal} {Phys.Rev.Lett.}}\ }%
  \textbf{\bibinfo {volume} {104}},\ \bibinfo {pages} {091801} (\bibinfo {year}
  {2010}),\ \Eprint{http://arxiv.org/abs/0912.1057}{arXiv:0912.1057 [hep-ex]}%
  \bibAnnoteFile{NoStop}{Aaltonen:2009nr}%
\bibitem{lhcdilepton}%
  \BibitemOpen
  \bibfield{author}{%
  \bibinfo {author} {\bibnamefont{{The CMS Collaboration}}},\ }%
  \bibinfo {howpublished}
  {\url{https://twiki.cern.ch/twiki/bin/view/CMSPublic/PhysicsResultsSUS10004},
  and see Colin Bernet's talk at Rencontres de Moriond EW 2011} (\bibinfo
  {year} {2011})%
  \bibAnnoteFile{NoStop}{lhcdilepton}%
\bibitem{Cheung:2009ch}%
  \BibitemOpen
  \bibfield{author}{%
  \bibinfo {author} {\bibfnamefont{K.}~\bibnamefont{Cheung}}, \bibinfo {author}
  {\bibfnamefont{W.-Y.}\ \bibnamefont{Keung}},\ and\ \bibinfo {author}
  {\bibfnamefont{T.-C.}\ \bibnamefont{Yuan}},\ }%
  \bibfield{journal}{%
  \Doi{10.1016/j.physletb.2009.11.015}{\bibinfo {journal} {Phys. Lett.}}\ }%
  \textbf{\bibinfo {volume} {B682}},\ \bibinfo {pages} {287} (\bibinfo {year}
  {2009}),\ \Eprint{http://arxiv.org/abs/0908.2589}{arXiv:0908.2589 [hep-ph]}%
  \bibAnnoteFile{NoStop}{Cheung:2009ch}%
%%CITATION = 0908.2589;%%
\bibitem{Cao:2010zb}%
  \BibitemOpen
  \bibfield{author}{%
  \bibinfo {author} {\bibfnamefont{Q.-H.}\ \bibnamefont{Cao}}, \bibinfo
  {author} {\bibfnamefont{D.}~\bibnamefont{McKeen}}, \bibinfo {author}
  {\bibfnamefont{J.~L.}\ \bibnamefont{Rosner}}, \bibinfo {author}
  {\bibfnamefont{G.}~\bibnamefont{Shaughnessy}},\ and\ \bibinfo {author}
  {\bibfnamefont{C.~E.~M.}\ \bibnamefont{Wagner}},\ }%
  \bibfield{journal}{%
  \Doi{10.1103/PhysRevD.81.114004}{\bibinfo {journal} {Phys. Rev.}}\ }%
  \textbf{\bibinfo {volume} {D81}},\ \bibinfo {pages} {114004} (\bibinfo {year}
  {2010}),\ \Eprint{http://arxiv.org/abs/1003.3461}{arXiv:1003.3461 [hep-ph]}%
  \bibAnnoteFile{NoStop}{Cao:2010zb}%
%%CITATION = 1003.3461;%%
\bibitem{Jung:2009jz}%
  \BibitemOpen
  \bibfield{author}{%
  \bibinfo {author} {\bibfnamefont{S.}~\bibnamefont{Jung}}, \bibinfo {author}
  {\bibfnamefont{H.}~\bibnamefont{Murayama}}, \bibinfo {author}
  {\bibfnamefont{A.}~\bibnamefont{Pierce}},\ and\ \bibinfo {author}
  {\bibfnamefont{J.~D.}\ \bibnamefont{Wells}},\ }%
  \bibfield{journal}{%
  \Doi{10.1103/PhysRevD.81.015004}{\bibinfo {journal} {Phys. Rev.}}\ }%
  \textbf{\bibinfo {volume} {D81}},\ \bibinfo {pages} {015004} (\bibinfo {year}
  {2010}),\ \Eprint{http://arxiv.org/abs/0907.4112}{arXiv:0907.4112 [hep-ph]}%
  \bibAnnoteFile{NoStop}{Jung:2009jz}%
%%CITATION = 0907.4112;%%
\bibitem{Gresham:2011pa}%
  \BibitemOpen
  \bibfield{author}{%
  \bibinfo {author} {\bibfnamefont{M.~I.}\ \bibnamefont{Gresham}}, \bibinfo
  {author} {\bibfnamefont{I.-W.}\ \bibnamefont{Kim}},\ and\ \bibinfo {author}
  {\bibfnamefont{K.~M.}\ \bibnamefont{Zurek}}}%
   (\bibinfo {year} {2011}),\
  \Eprint{http://arxiv.org/abs/1103.3501}{arXiv:1103.3501 [hep-ph]}%
  \bibAnnoteFile{NoStop}{Gresham:2011pa}%
\bibitem{cdftprime}%
  \BibitemOpen
  \bibfield{author}{%
  \bibinfo {author} {\bibnamefont{{The CDF Collaboration}}},\ }%
  \bibinfo {howpublished} {CDF Note 10110} (\bibinfo {year} {2010})%
  \bibAnnoteFile{NoStop}{cdftprime}%
\bibitem{Aaltonen:2009iz}%
  \BibitemOpen
  \bibfield{author}{%
  \bibinfo {author} {\bibfnamefont{T.}~\bibnamefont{Aaltonen}} \emph{et~al.}
  (\bibinfo {collaboration} {CDF Collaboration}),\ }%
  \bibfield{journal}{%
  \Doi{10.1103/PhysRevLett.102.222003}{\bibinfo {journal} {Phys.Rev.Lett.}}\ }%
  \textbf{\bibinfo {volume} {102}},\ \bibinfo {pages} {222003} (\bibinfo {year}
  {2009}),\ \Eprint{http://arxiv.org/abs/0903.2850}{arXiv:0903.2850 [hep-ex]}%
  \bibAnnoteFile{NoStop}{Aaltonen:2009iz}%
\bibitem{Finkbeiner:2007kk}%
  \BibitemOpen
  \bibfield{author}{%
  \bibinfo {author} {\bibfnamefont{D.~P.}\ \bibnamefont{Finkbeiner}}\ and\
  \bibinfo {author} {\bibfnamefont{N.}~\bibnamefont{Weiner}},\ }%
  \bibfield{journal}{%
  \Doi{10.1103/PhysRevD.76.083519}{\bibinfo {journal} {Phys. Rev.}}\ }%
  \textbf{\bibinfo {volume} {D76}},\ \bibinfo {pages} {083519} (\bibinfo {year}
  {2007}),\
  \Eprint{http://arxiv.org/abs/astro-ph/0702587}{arXiv:astro-ph/0702587}%
  \bibAnnoteFile{NoStop}{Finkbeiner:2007kk}%
%%CITATION = ASTRO-PH/0702587;%%
\bibitem{Pospelov:2007mp}%
  \BibitemOpen
  \bibfield{author}{%
  \bibinfo {author} {\bibfnamefont{M.}~\bibnamefont{Pospelov}}, \bibinfo
  {author} {\bibfnamefont{A.}~\bibnamefont{Ritz}},\ and\ \bibinfo {author}
  {\bibfnamefont{M.~B.}\ \bibnamefont{Voloshin}},\ }%
  \bibfield{journal}{%
  \Doi{10.1016/j.physletb.2008.02.052}{\bibinfo {journal} {Phys. Lett.}}\ }%
  \textbf{\bibinfo {volume} {B662}},\ \bibinfo {pages} {53} (\bibinfo {year}
  {2008}),\ \Eprint{http://arxiv.org/abs/0711.4866}{arXiv:0711.4866 [hep-ph]}%
  \bibAnnoteFile{NoStop}{Pospelov:2007mp}%
%%CITATION = 0711.4866;%%
\bibitem{Cheung:2009qd}%
  \BibitemOpen
  \bibfield{author}{%
  \bibinfo {author} {\bibfnamefont{C.}~\bibnamefont{Cheung}}, \bibinfo {author}
  {\bibfnamefont{J.~T.}\ \bibnamefont{Ruderman}}, \bibinfo {author}
  {\bibfnamefont{L.-T.}\ \bibnamefont{Wang}},\ and\ \bibinfo {author}
  {\bibfnamefont{I.}~\bibnamefont{Yavin}},\ }%
  \bibfield{journal}{%
  \Doi{10.1103/PhysRevD.80.035008}{\bibinfo {journal} {Phys. Rev.}}\ }%
  \textbf{\bibinfo {volume} {D80}},\ \bibinfo {pages} {035008} (\bibinfo {year}
  {2009}),\ \Eprint{http://arxiv.org/abs/0902.3246}{arXiv:0902.3246 [hep-ph]}%
  \bibAnnoteFile{NoStop}{Cheung:2009qd}%
%%CITATION = 0902.3246;%%
\bibitem{Feng:2011vu}%
  \BibitemOpen
  \bibfield{author}{%
  \bibinfo {author} {\bibfnamefont{J.~L.}\ \bibnamefont{Feng}}, \bibinfo
  {author} {\bibfnamefont{J.}~\bibnamefont{Kumar}}, \bibinfo {author}
  {\bibfnamefont{D.}~\bibnamefont{Marfatia}},\ and\ \bibinfo {author}
  {\bibfnamefont{D.}~\bibnamefont{Sanford}}}%
   (\bibinfo {year} {2011}),\
  \Eprint{http://arxiv.org/abs/1102.4331}{arXiv:1102.4331 [hep-ph]}%
  \bibAnnoteFile{NoStop}{Feng:2011vu}%
%%CITATION = 1102.4331;%%
\bibitem{Cohen:2010gj}%
  \BibitemOpen
  \bibfield{author}{%
  \bibinfo {author} {\bibfnamefont{T.}~\bibnamefont{Cohen}}, \bibinfo {author}
  {\bibfnamefont{D.~J.}\ \bibnamefont{Phalen}},\ and\ \bibinfo {author}
  {\bibfnamefont{A.}~\bibnamefont{Pierce}},\ }%
  \bibfield{journal}{%
  \Doi{10.1103/PhysRevD.81.116001}{\bibinfo {journal} {Phys. Rev.}}\ }%
  \textbf{\bibinfo {volume} {D81}},\ \bibinfo {pages} {116001} (\bibinfo {year}
  {2010}),\ \Eprint{http://arxiv.org/abs/1001.3408}{arXiv:1001.3408 [hep-ph]}%
  \bibAnnoteFile{NoStop}{Cohen:2010gj}%
%%CITATION = 1001.3408;%%
\bibitem{Goldberg:1983nd}%
  \BibitemOpen
  \bibfield{author}{%
  \bibinfo {author} {\bibfnamefont{H.}~\bibnamefont{Goldberg}},\ }%
  \bibfield{journal}{%
  \Doi{10.1103/PhysRevLett.50.1419}{\bibinfo {journal} {Phys. Rev. Lett.}}\ }%
  \textbf{\bibinfo {volume} {50}},\ \bibinfo {pages} {1419} (\bibinfo {year}
  {1983})%
  \bibAnnoteFile{NoStop}{Goldberg:1983nd}%
%%CITATION = PRLTA,50,1419;%%
\bibitem{Cirelli:2005gh}%
  \BibitemOpen
  \bibfield{author}{%
  \bibinfo {author} {\bibfnamefont{M.}~\bibnamefont{Cirelli}} \emph{et~al.},\
  }%
  \bibfield{journal}{%
  \Doi{10.1016/j.nuclphysb.2005.08.017}{\bibinfo {journal} {Nucl. Phys.}}\ }%
  \textbf{\bibinfo {volume} {B727}},\ \bibinfo {pages} {99} (\bibinfo {year}
  {2005}),\ \Eprint{http://arxiv.org/abs/hep-ph/0506298}{arXiv:hep-ph/0506298}%
  \bibAnnoteFile{NoStop}{Cirelli:2005gh}%
%%CITATION = HEP-PH/0506298;%%
\bibitem{Cholis:2008vb}%
  \BibitemOpen
  \bibfield{author}{%
  \bibinfo {author} {\bibfnamefont{I.}~\bibnamefont{Cholis}}, \bibinfo {author}
  {\bibfnamefont{L.}~\bibnamefont{Goodenough}},\ and\ \bibinfo {author}
  {\bibfnamefont{N.}~\bibnamefont{Weiner}},\ }%
  \bibfield{journal}{%
  \Doi{10.1103/PhysRevD.79.123505}{\bibinfo {journal} {Phys. Rev.}}\ }%
  \textbf{\bibinfo {volume} {D79}},\ \bibinfo {pages} {123505} (\bibinfo {year}
  {2009}),\ \Eprint{http://arxiv.org/abs/0802.2922}{arXiv:0802.2922
  [astro-ph]}%
  \bibAnnoteFile{NoStop}{Cholis:2008vb}%
%%CITATION = 0802.2922;%%
\bibitem{ArkaniHamed:2008qn}%
  \BibitemOpen
  \bibfield{author}{%
  \bibinfo {author} {\bibfnamefont{N.}~\bibnamefont{Arkani-Hamed}}, \bibinfo
  {author} {\bibfnamefont{D.~P.}\ \bibnamefont{Finkbeiner}}, \bibinfo {author}
  {\bibfnamefont{T.~R.}\ \bibnamefont{Slatyer}},\ and\ \bibinfo {author}
  {\bibfnamefont{N.}~\bibnamefont{Weiner}},\ }%
  \bibfield{journal}{%
  \Doi{10.1103/PhysRevD.79.015014}{\bibinfo {journal} {Phys. Rev.}}\ }%
  \textbf{\bibinfo {volume} {D79}},\ \bibinfo {pages} {015014} (\bibinfo {year}
  {2009}),\ \Eprint{http://arxiv.org/abs/0810.0713}{arXiv:0810.0713 [hep-ph]}%
  \bibAnnoteFile{NoStop}{ArkaniHamed:2008qn}%
%%CITATION = 0810.0713;%%
\bibitem{Pospelov:2008jd}%
  \BibitemOpen
  \bibfield{author}{%
  \bibinfo {author} {\bibfnamefont{M.}~\bibnamefont{Pospelov}}\ and\ \bibinfo
  {author} {\bibfnamefont{A.}~\bibnamefont{Ritz}},\ }%
  \bibfield{journal}{%
  \Doi{10.1016/j.physletb.2008.12.012}{\bibinfo {journal} {Phys. Lett.}}\ }%
  \textbf{\bibinfo {volume} {B671}},\ \bibinfo {pages} {391} (\bibinfo {year}
  {2009}),\ \Eprint{http://arxiv.org/abs/0810.1502}{arXiv:0810.1502 [hep-ph]}%
  \bibAnnoteFile{NoStop}{Pospelov:2008jd}%
%%CITATION = 0810.1502;%%
\bibitem{Fox:2008kb}%
  \BibitemOpen
  \bibfield{author}{%
  \bibinfo {author} {\bibfnamefont{P.~J.}\ \bibnamefont{Fox}}\ and\ \bibinfo
  {author} {\bibfnamefont{E.}~\bibnamefont{Poppitz}},\ }%
  \bibfield{journal}{%
  \Doi{10.1103/PhysRevD.79.083528}{\bibinfo {journal} {Phys. Rev.}}\ }%
  \textbf{\bibinfo {volume} {D79}},\ \bibinfo {pages} {083528} (\bibinfo {year}
  {2009}),\ \Eprint{http://arxiv.org/abs/0811.0399}{arXiv:0811.0399 [hep-ph]}%
  \bibAnnoteFile{NoStop}{Fox:2008kb}%
%%CITATION = 0811.0399;%%
\bibitem{Note2}%
  \BibitemOpen
  \bibinfo {note} {Or equivalently, a much smaller leptonic BR}%
  \bibAnnoteFile{NoStop}{Note2}%
\end{thebibliography}%

\end{document}